\documentclass[reqno]{article}
\usepackage{psfig}
\usepackage{amsmath}
\begin{document}

\title{Core collapse supernovae: magnetorotational explosion.}

\bigskip
\author{G.S.Bisnovatyi-Kogan$^{1,2}$, S.G.Moiseenko$^1$ and N.V.Ardeljan$^3$\\
\small\it $^1$IKI RAS, Profsoyuznaya 84/32, Moscow 117997, Russia\\[-1.mm]
 \small\it $^2$JINR, Dubna, Russia \\[-1.mm]
\small\it $^3$Department of Computational Mathematics and
Cybernetics, Moscow State University,  \\[-1.mm]
\small\it Vorobjevy Gory, Moscow B-234, Russia
  \\[-1.mm]
\small\it email: gkogan@iki.rssi.ru}
\date{ }
\maketitle
\bigskip

\maketitle
\bigskip

\begin{abstract}

 Core-collapse supernovae are connected with formation of neutron
 stars. Part of the gravitation energy is transformed into the energy of
 the  explosion, observed in SN II, SN Ib,c type supernovae. The mechanism
 of transformation is not simple, because the overwhelming majority of the
 energy is going into weakly interacting neutrino. The attempts to use this
 energy for the explosion were not successful during about 40 years of
 investigation.
   We consider the explosion mechanism in which the source of energy is the
 rotation, and magnetic field serves for the transformation of the rotation
 energy into the energy of explosion. 2-D MHD simulations of this
 mechanism were performed. After the collapse the core consists of a
 rapidly rotating proto-neutron star with a differentially rotating
 envelope. The  toroidal part of the magnetic energy generated by the
 differential rotation grows as quadratic function with time at the
 initial stage of the
 evolution of the magnetic field. The linear growth of the toroidal
 magnetic field is terminated by the development of magnetohydrodynamic
 instability, when the twisted toroidal component strongly exceeds the poloidal field,
 leading to a drastic acceleration in the growth of magnetic energy. At the moment
 when the magnetic pressure becomes comparable to the gas pressure at the
 periphery of the proto-neutron star the MHD compression wave appears and
 goes through the envelope of the collapsed core. It transforms into the
 fast MHD shock and produces a supernova
 explosion. Our simulations give the energy of the explosion $0.6\cdot
 10^{51}$ ergs.
 The amount of the mass ejected by the explosion is $\sim 0.14M_\odot$.
 The implicit numerical method, based on the Lagrangian triangular grid of
 variable structure, was used for the simulations.

\end{abstract}

\section{Introduction}

{\bf Content}.

1. Presupernovae.

2. Magnetorotational mechanism of explosion:
1-D calculations.

3. 2-D MHD: Numerical method.

4. Core collapse and formation of rapidly rotating neutron star.

5. Magnetorotational supernova explosion.

6. Development of magnetorotational instability.

7. Jet formation in magnetorotational SN explosion.

8. Mirror symmetry breaking: Rapidly moving pulsars.  One-side jets.

\begin{figure}
\label{star}
\centerline{\psfig{figure=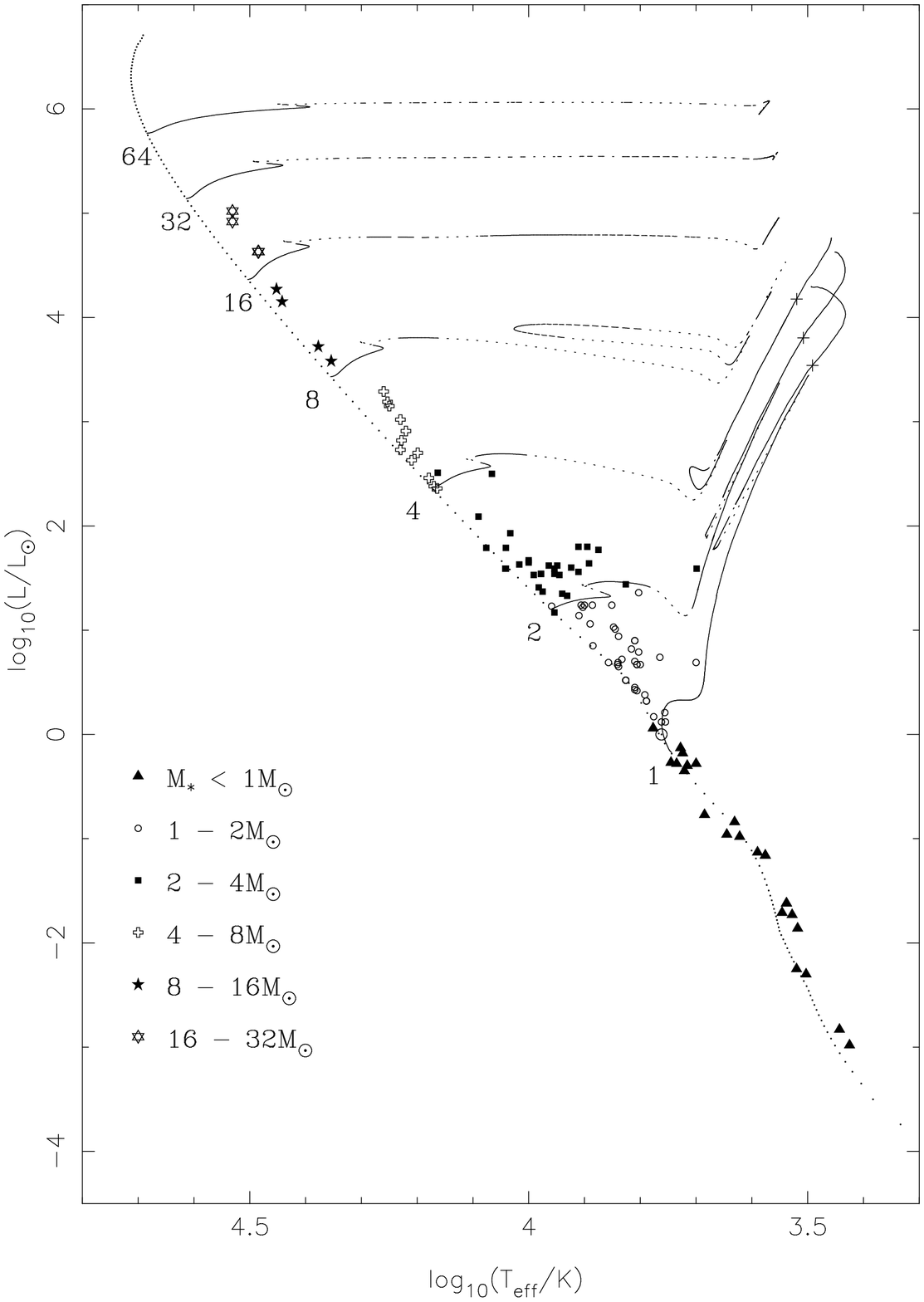,width=3in}
           \psfig{figure=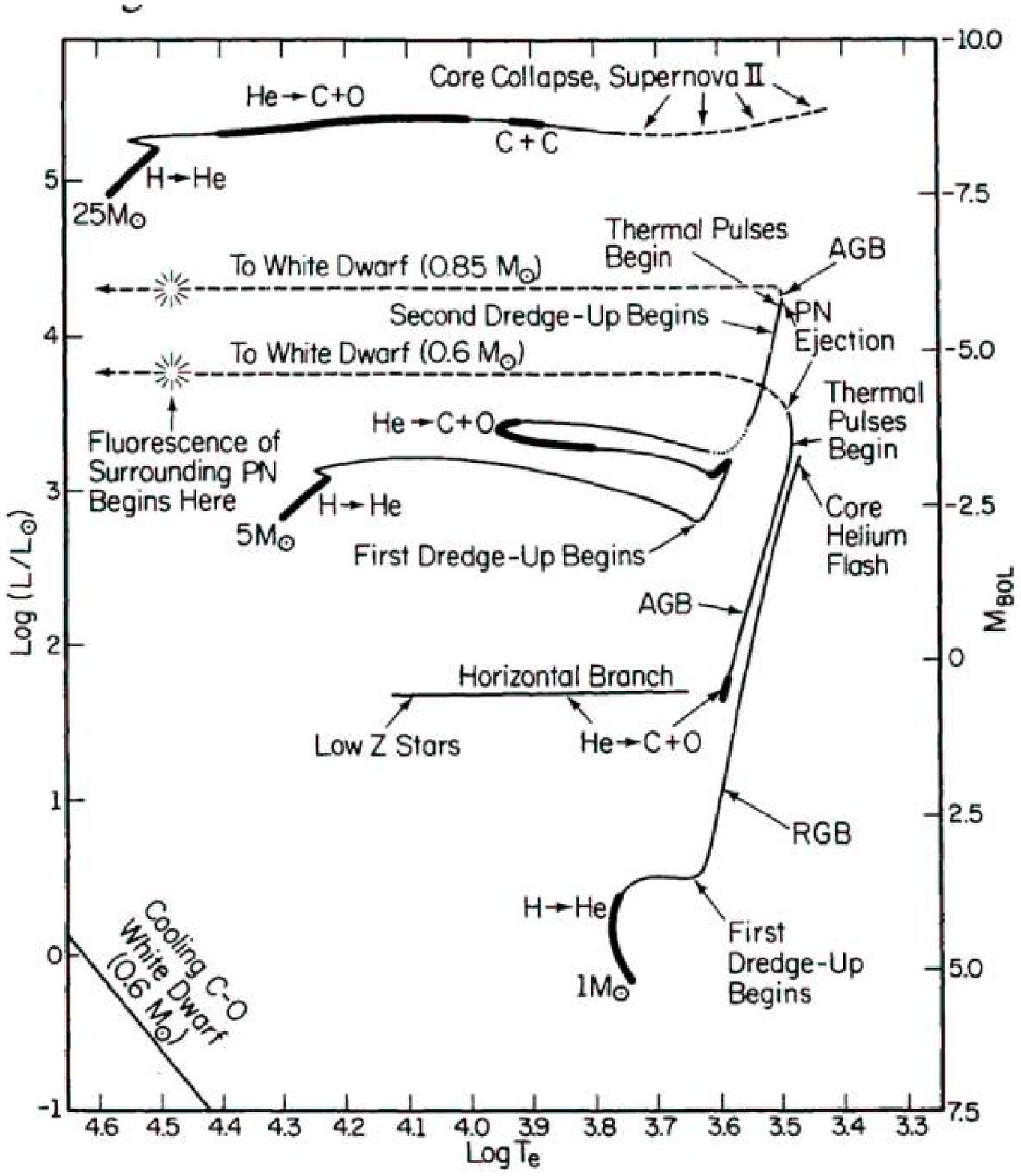,width=3in}}
\caption
{{\small
 {\bf Left.} Hertzsprung--Russell diagram of the ZAMS and several evolutionary tracks.
 The masses of the models are indicated at the starts of the tracks, in
solar units. The solid portions of the tracks indicate where
evolution is on a relatively slow nuclear time-scale, the dotted parts show
evolution on a thermal time-scale, and the dashed parts show
an intermediate time-scale. The different symbols indicate the positions of
binary components with well-determined masses, radii and
luminosities. The position of the Sun is indicated by a solar symbol ${\odot}$.
The crosses (+) show the position on the giant branches of
the 1-, 2- and 4-M$_{\odot}$ models where the binding energy of the envelope becomes
 positive, from \cite{pols94}.
\newline
 \qquad{\bf Right.} Coarse evolutionary tracks for stars with $M_i=1,5,25\,M_\odot$.
Heavy portions represent principal burning phases in the core. For
$M_i<2.3\,M_\odot$, CHF (core helium flash) occurs after which
quiescent $^4$He burning begins. After $^4$He has been exhausted
in the core, the star arrives on the AGB (asymptotic giant branch).
When the core with no helium in it reaches the mass
$\sim 0.53\,M_\odot$, TF (thermal flashes in $^4$He shell) start.
An AGB star loses mass, and this process terminates with a rapid
ejection of the residual hydrogen envelope in the form of PN
(planetary nebula). CO core with $M_f\approx 0.6\,M_\odot$
transforms into white dwarf. More massive AGB and post-AGB stars
with $M_i\le 9\,M_\odot$ evolve in a similar way, $M_f$
increases with increasing $M_i$ and equals $1.08\,M_\odot$ at
$M_i=8.8\,M_\odot$. The symbol ~{\small$\bigcirc$}~ indicates the
onset of
planetary nebula luminescence, when $T_{ef}$ of the star reaches
$3\cdot 10^4$~K and the gas ionization in PN begins, from \cite{iben85}.}}
\end{figure}

\noindent Supernova is one of the most powerful explosions in the
Universe, producing the energy (radiation and kinetic) about
$10^{51}$ erg. The explosion happens at the end of the evolution
of massive stars, with initial mass more than $\sim$8 Solar
masses. Hertzsprung-Russell diagram of the zero age main sequence,
and several evolutional tracks are represented in left side of
Fig.\ref{star} from \cite{pols94}. Tracks in HR diagram of  a
representative selection of stars from the main sequence till the
end of the evolution are given in the right side of Fig.\ref{star}
from \cite{iben85}.

The explosion mechanisms are:

1. Thermonuclear explosion of C-O degenerate core (SN Ia)

2. Core collapse and formation of a neutron star, with a gravitational energy

\quad release $\sim 6 \cdot 10^{53}$ erg, carried away by neutrino (SN II, SN Ib,c).

\noindent
The transformation of the neutrino energy into kinetic one is very problematic,
and was not obtained firmly. Magnetorotational explosion (MRE), in the
core collapse SN, is based on the
transformation of the rotational energy of the neutron star into the
explosion energy by means of the magnetic field.

Most of supernova explosions and ejections are not spherically
symmetrical. A lot of stars are rotating and have magnetic fields.
Often we can see one-side ejections. One of the main difficulties
for core collapse supernova explosions is, how to transform
any kind of energy of the star into the explosion energy.
The magnetorotational mechanism suggested in \cite{bk70},
transforms the rotational energy of the star
into the explosion energy with a high efficiency of about 10\%.
It is acting in presence of the differential
rotation when the rotational energy
can be transformed to the explosion energy by the magnetic field action.

\section{1-D calculations}

\begin{figure}
\centerline{\psfig{figure=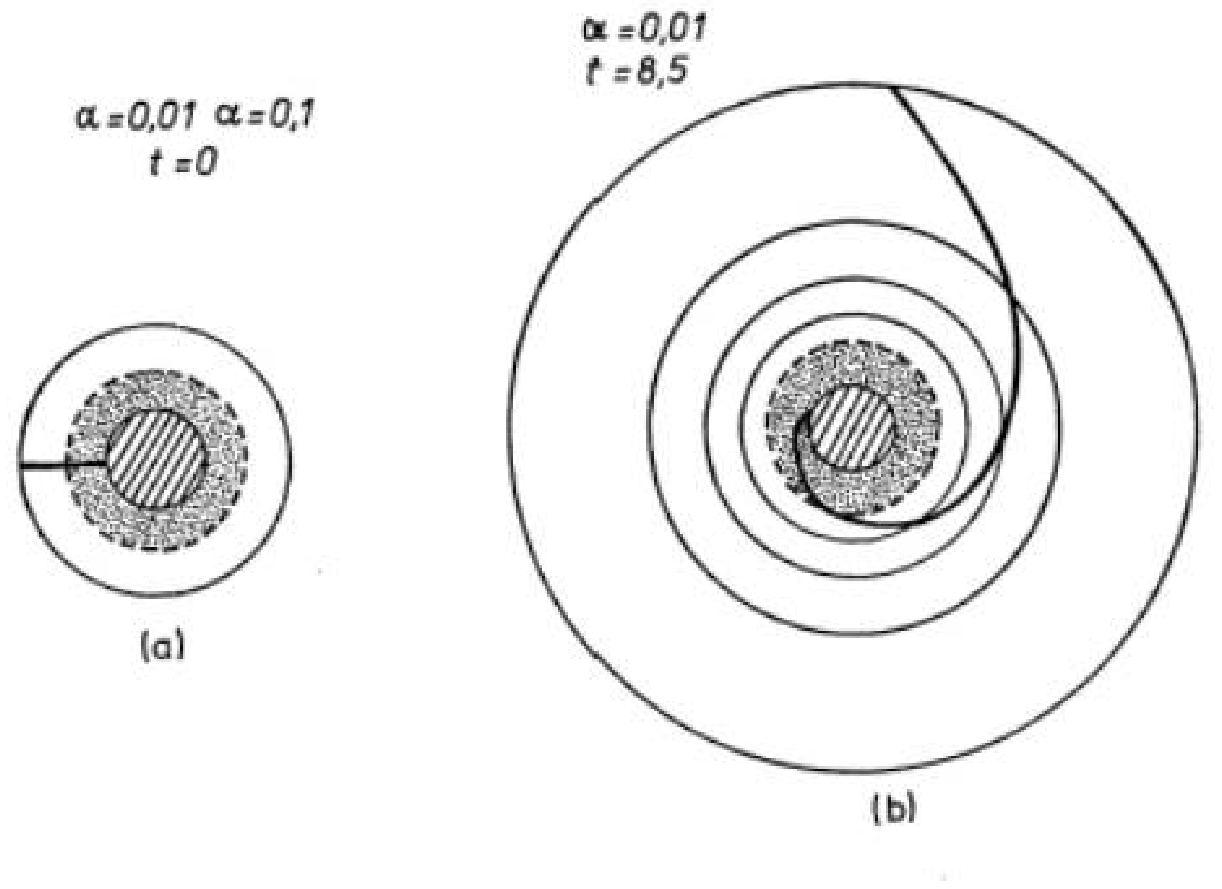,width=3.2in}
\psfig{figure=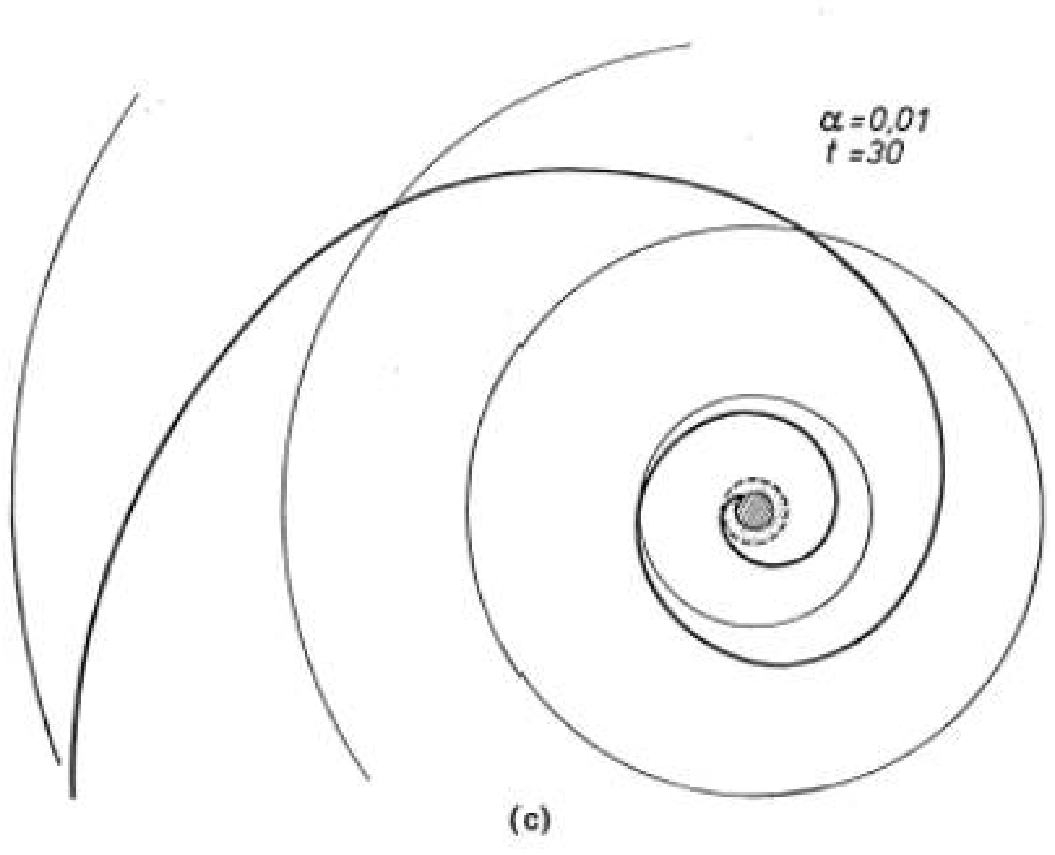,width=3.2in}}
\caption
{{\small
The configuration of the magnetic field line in the subsequent moments of
time for $\alpha=0.01$. The dashed region is the central incompressible core of
the neutron star; the pointed region is a contracting part of the envelope.
In the right figure (c) the scale is diminished to a half in comparison
with (a) and (b), from \cite{bkps76}.
}}
\label{alph001}
\end{figure}

\begin{figure}
\centerline{\psfig{figure=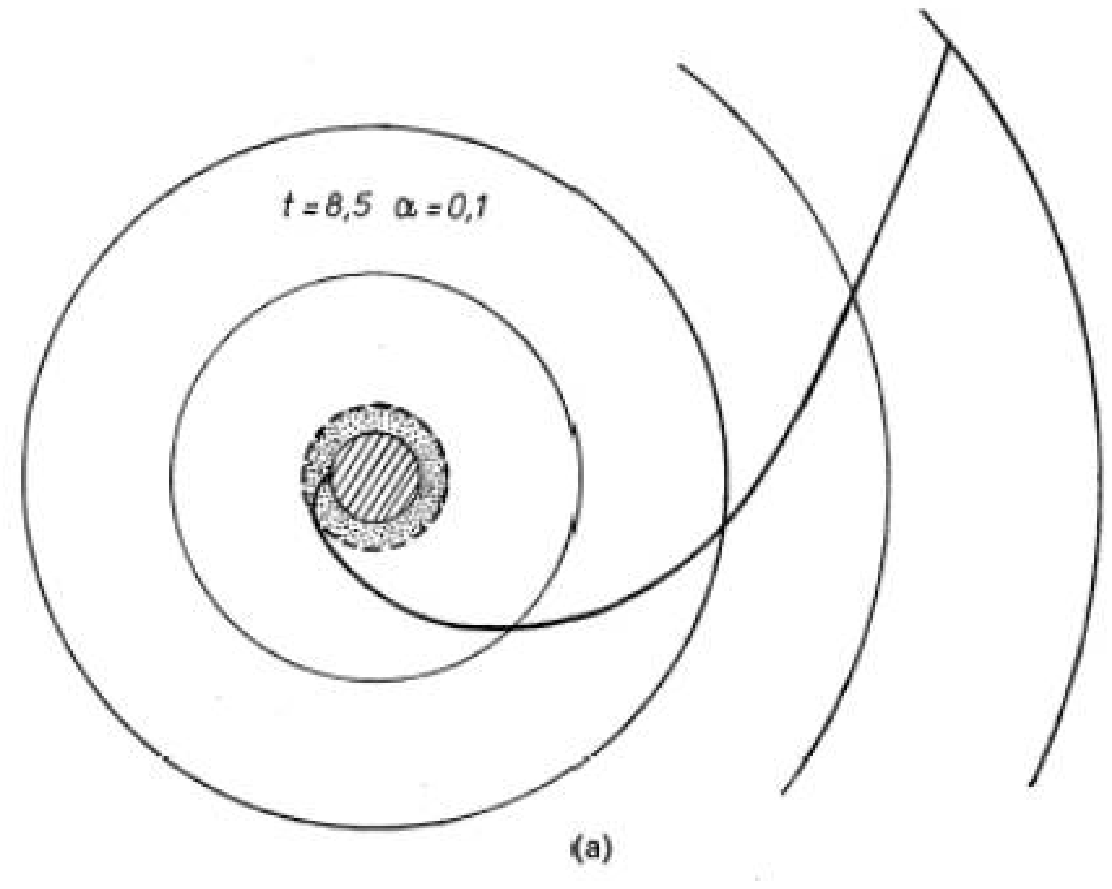,width=3.2in}
\psfig{figure=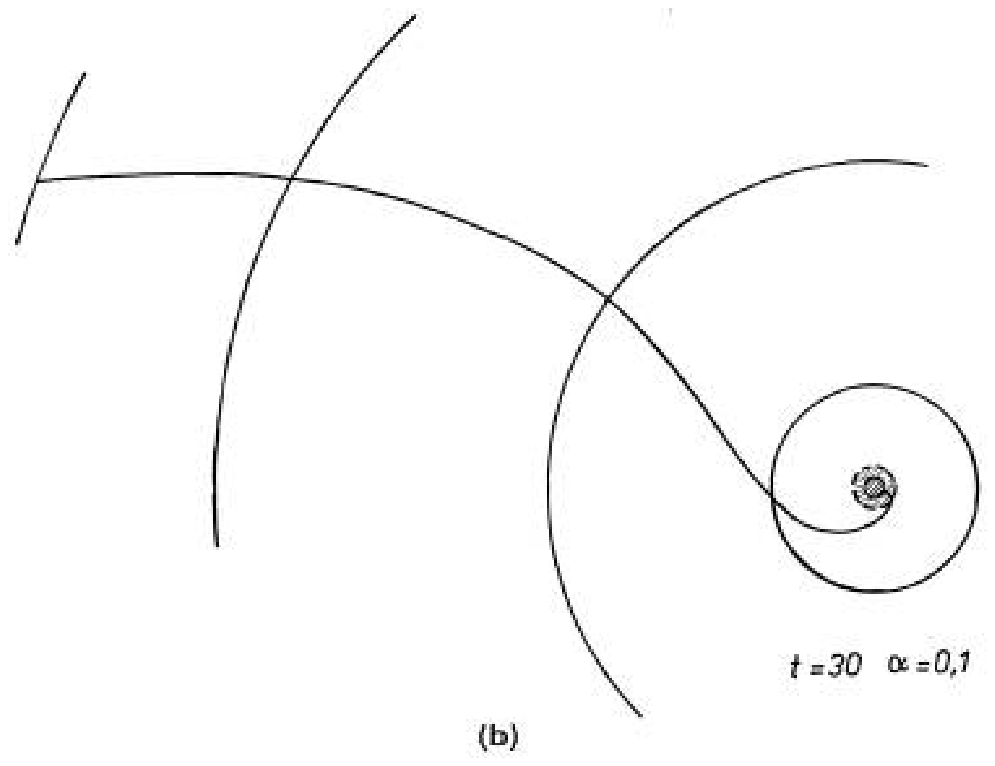,width=3.2in}}
\caption
{{\small
The same picture as in Fig.\ref{alph001}, but for $\alpha=0.1$.
The initial picture coincides with Fig.\ref{alph001}. In the
figure (b) the scale is diminished four times, from \cite{bkps76}.}}
\label{alph01}
\end{figure}

1-D calculations of the magnetorotational explosion (MRE) have been performed
first in \cite{bkps76}.
The important parameter of the problem is the ratio of the magnetic and
gravitational energies at the initial moment, when differential rotation starts
to twist and to amplify the toroidal component, $\alpha=\frac{E_{mag0}}{E_{grav0}}$.
In \cite{bkps76} rather large values of $\alpha=0.1$ and 0.01 had been considered,
see Figs.\ref{alph001},\ref{alph01}.

\begin{figure}
\centerline{\psfig{figure=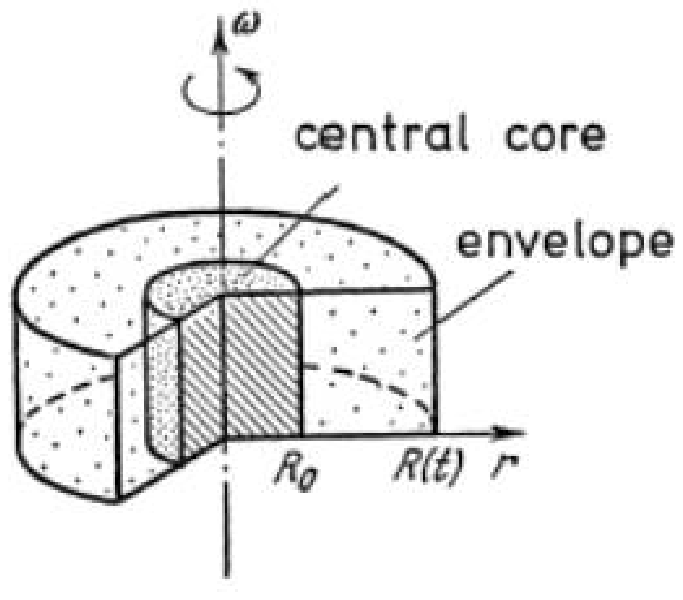,width=2.7in}
\psfig{figure=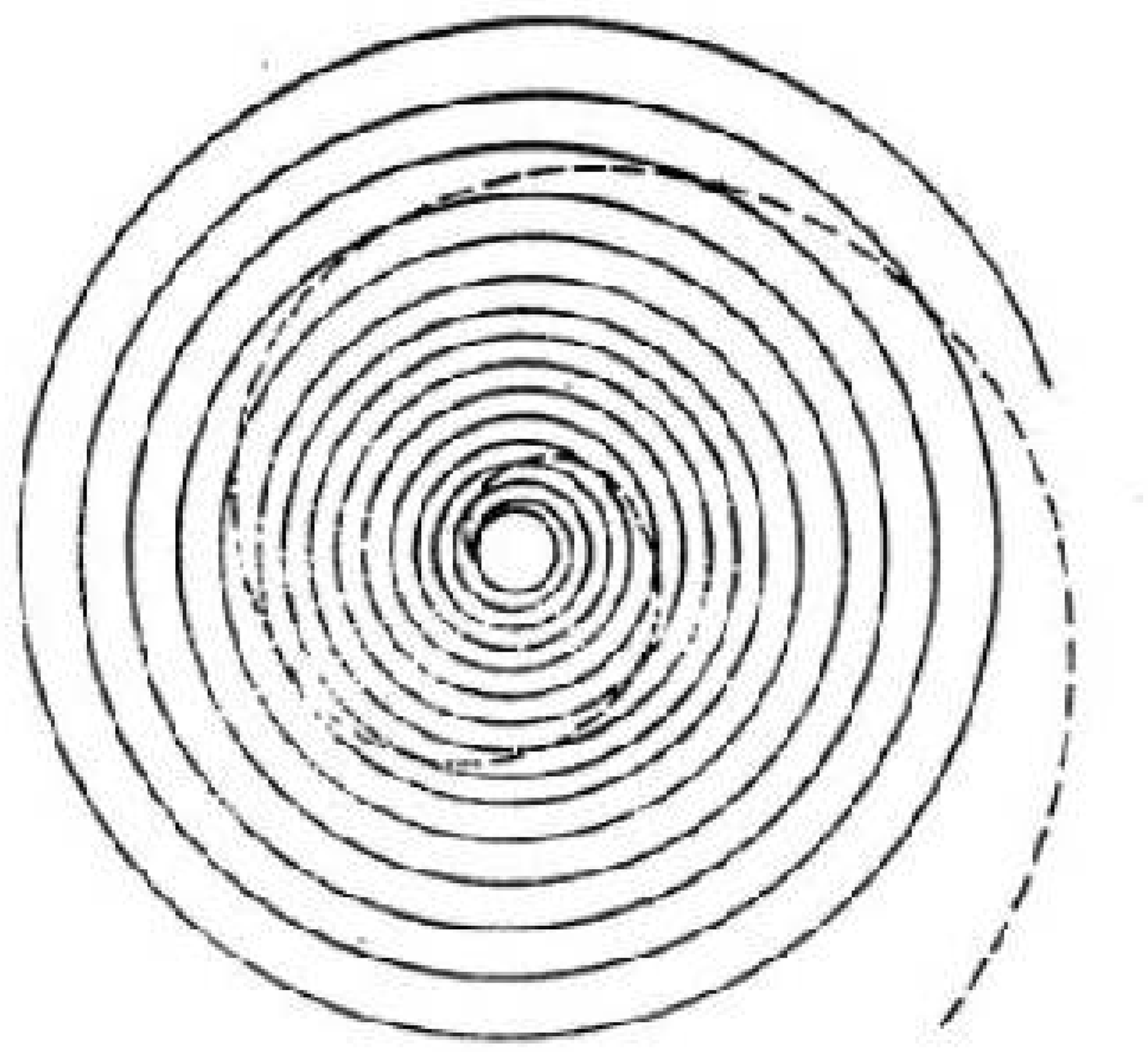,width=2.7in}}
\caption
{{\small {\bf left} A schematic picture of the initial state for the unit length of the
cylinder, from \cite{bkps76}.
\newline
{\bf right} Shape of the field lines in the region near the core at the time
$t=7/\sqrt{\alpha}$,  for $\alpha=0.01$ (dashed line), and $\alpha = 10^{-4}$
(solid line), from \cite{ard79}.}}
\label{field79}
\end{figure}
\begin{figure}
\centerline{\psfig{figure=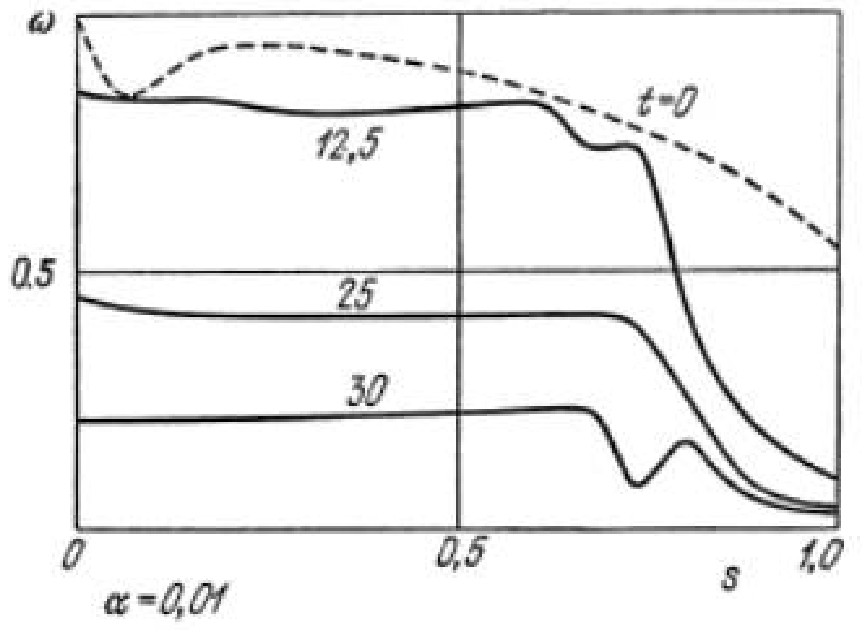,width=3.2in}
\psfig{figure=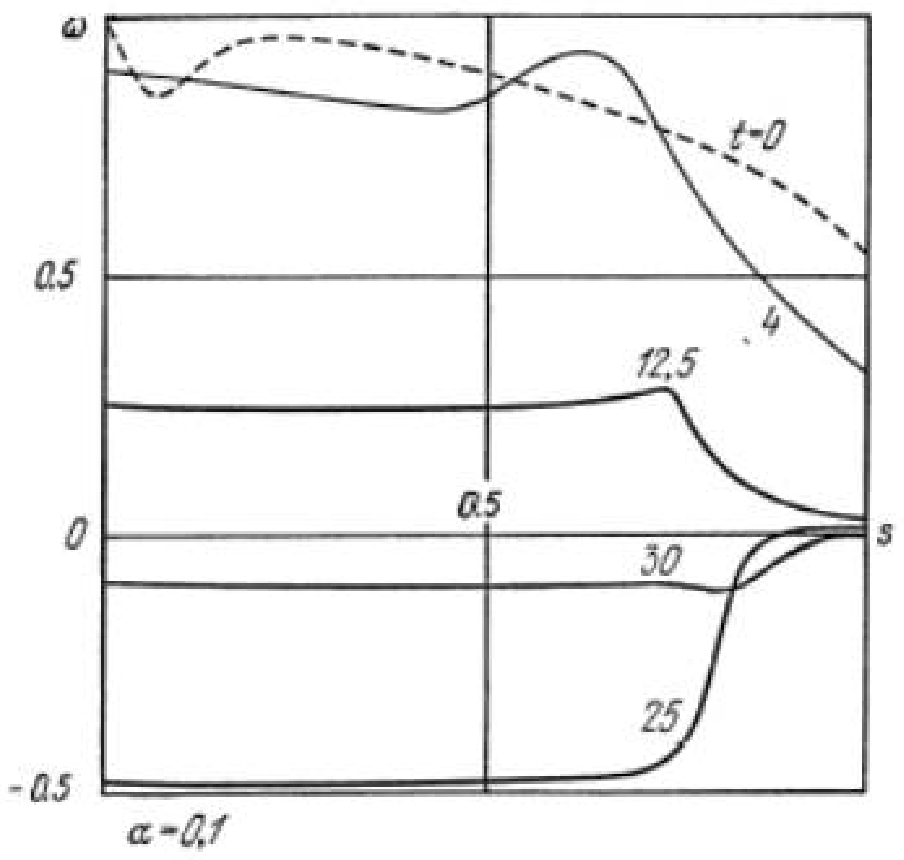,width=3.2in}}
\caption
{{\small
The dependence of the angular velocity $\omega$ on the mass coordinate $s$ at
later time for two values of the parameter $\alpha$. The division of the
envelope in two parts is seen: the uniformly rotating central part, and the
ejected outer part, in which the rotation became slower. The change of the
direction of rotation is seen on the right figure, from \cite{bkps76}.}}
\label{angvel}
\end{figure}
\begin{figure}
\centerline{\psfig{figure=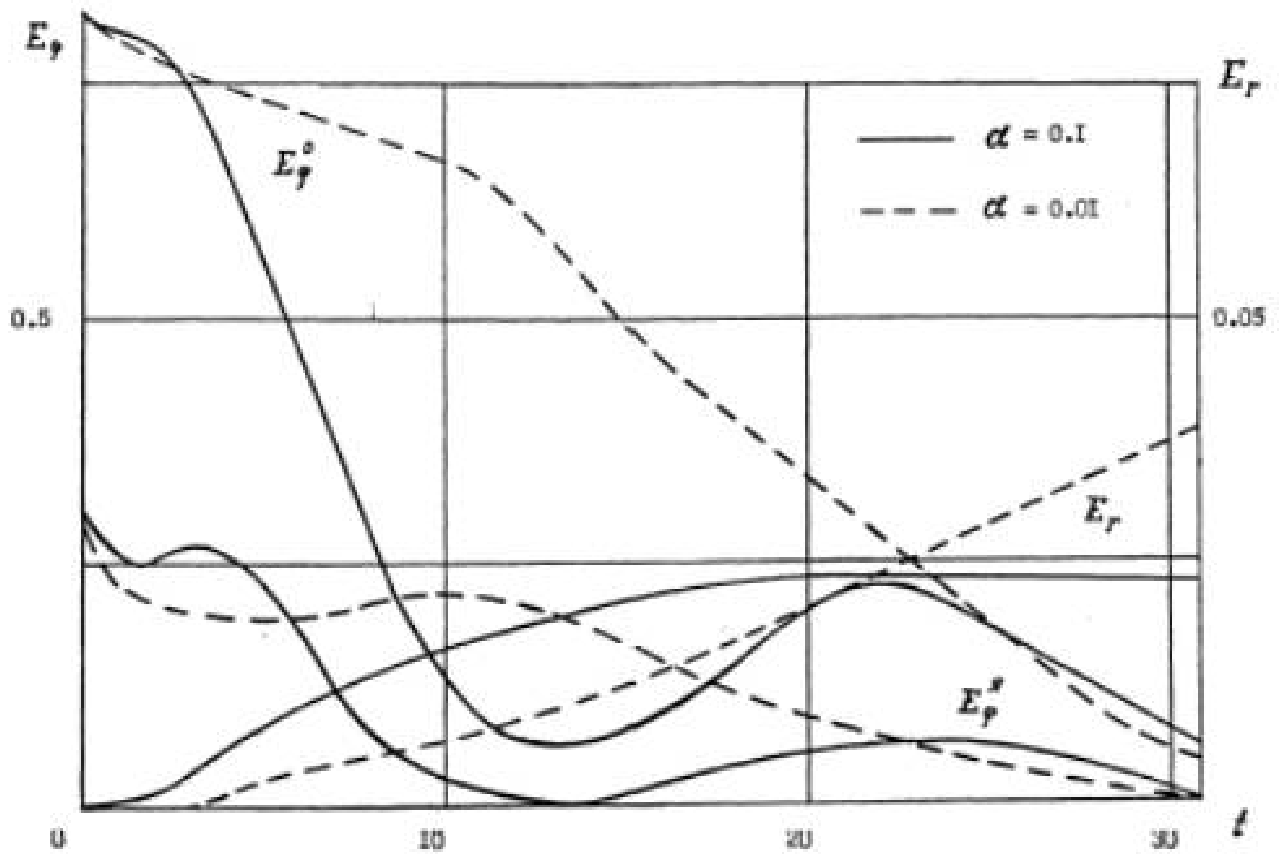,width=5in}}
\caption
{{\small
The dependence of the kinetic energy on time for two values of $\alpha$:
the rotational energy of the core $E_{\phi}^n$, envelope $E_{\phi}^0$,
the radial energy of the envelope $E_r$, from \cite{bkps76}.}}
\label{en76}
\end{figure}

1-D calculations of MRE for lower magnetic fields, at
$\alpha=10^{-2},\, 10^{-4},\, 10^{-8}$, had been done in
\cite{ard79}, see Fig.\ref{field79}.
Angular velocity distribution at different time moments.
is represented in Fig.\ref{angvel} form \cite{bkps76}.
Variations of different types of kinetic energy during MRE are given in Fig.\ref{en76} from
\cite{bkps76}.

The main results of 1-D calculations are the following.
Magneto-rotational explosion (MRE) has an efficiency about 10\% of
rotational energy. For the neutron star with mass $\approx 1.2
M_\odot$ the ejected mass is about $0.1 M_\odot$, the
explosion energy is about $10^{51}$ erg.  Ejected mass and
explosion energy depend weekly on the parameter $\alpha$,
explosion time strongly depends on $\alpha$, Explosion time
$t_{expl} \sim 1/\sqrt{\alpha}$ \cite{ard79}. Small $\alpha$  is
difficult for numerical calculations with EXPLICIT numerical
schemes because of  the Courant restriction on the time step for a
stiff set of equations: small $\alpha$  determines a stiffness. In
2-D numerical an IMPLICIT numerical schemes should be used.

\section{The difference scheme for 2-D calculations}

\begin{figure}
\centerline{\psfig{figure=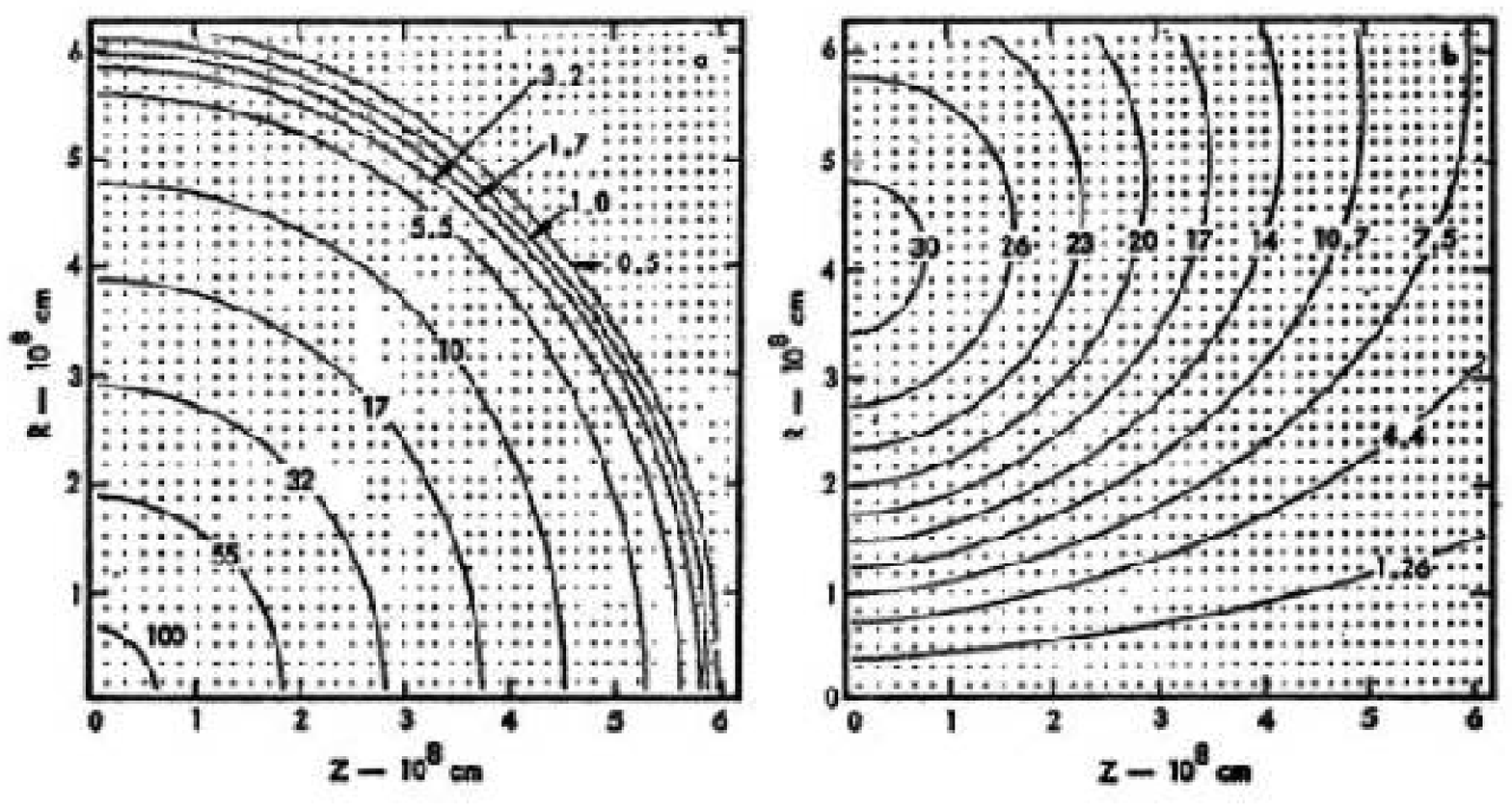,width=5in}}
\caption
{{\small
Isodensity contours in units of $10^6$ g/cm$^3$ {\bf (a)}, and
magnetic-flux contours parallel to Z-axis in units of
$10^{22}$ Gs/cm$^{-2}$ {\bf (b)},  at 0.72 sec, from \cite{lbwi70}.}}
\label{l1w70}
\end{figure}

\begin{figure}
\centerline{\psfig{figure=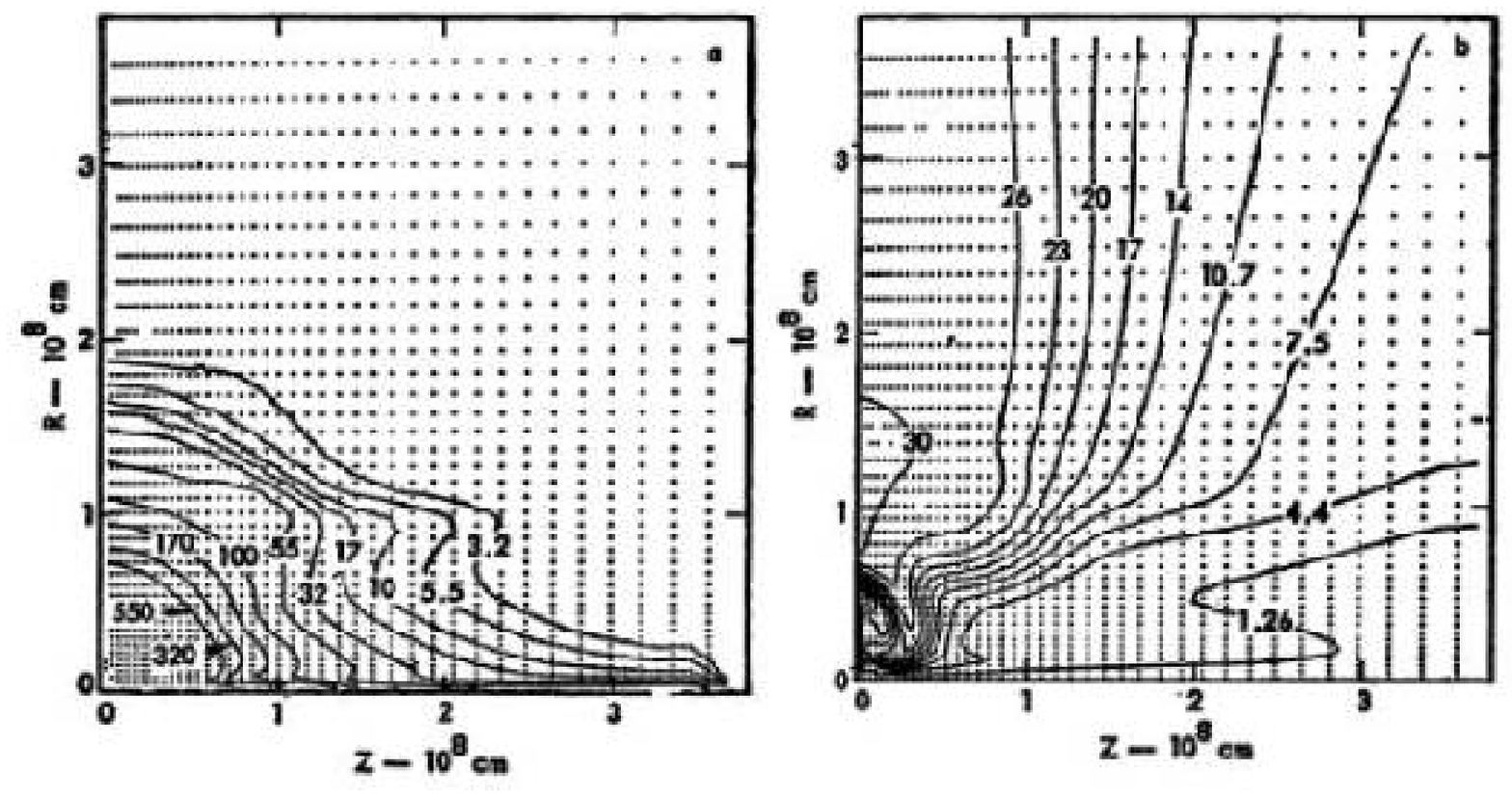,width=5in}}
\caption
{{\small
Isodensity contours in units of $10^6$ g/cm$^3$ {\bf (a)}, and
magnetic-flux contours parallel to Z-axis in units of
$10^{22}$ Gs/cm$^{-2}$ {\bf (b)}, at 2.67 sec, from \cite{lbwi70}.}}
\label{l2w70}
\end{figure}

First 2-D calculations, showing
jets in the collapse of rotating  magnetized star had been performed in
\cite{lbwi70}. Very large magnetic field was taken, so twisting of
magnetic field lines, expected in a realistic situation, was not important
for the explosion, see Figs.\ref{l1w70},\ref{l2w70}.

The difference scheme used in the present calculations is described in
\cite{ak95}, \cite{ard87}.
The scheme is based on the method of basic operators -
grid analogs of the main differential operators:

GRAD(scalar) (differential) ~ GRAD(scalar) (grid analog)

DIV(vector) (differential) ~ DIV(vector) (grid analog)

CURL(vector) (differential) ~ CURL(vector) (grid analog)

GRAD(vector) (differential) ~ GRAD(vector) (grid analog)

DIV(tensor) (differential) ~ DIV(tensor) (grid analog)

\begin{figure}
\centerline{\psfig{figure=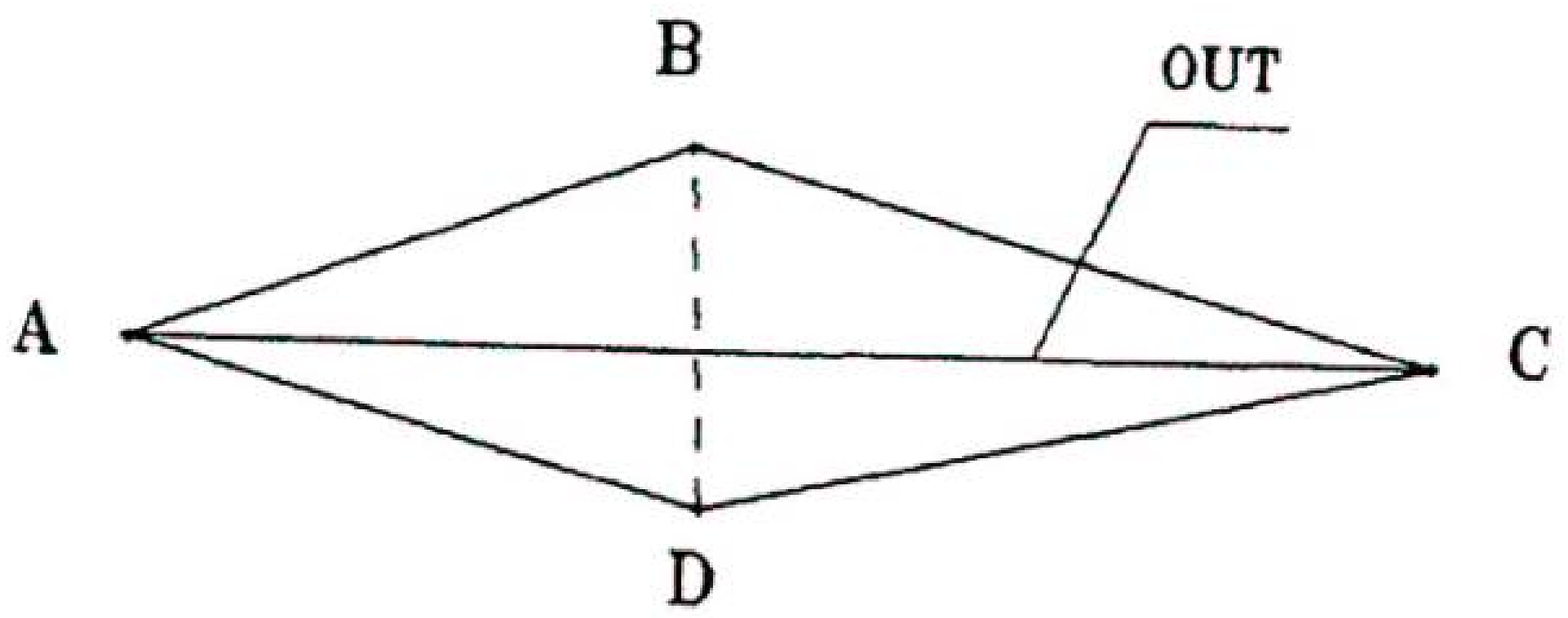,width=3in}
\psfig{figure=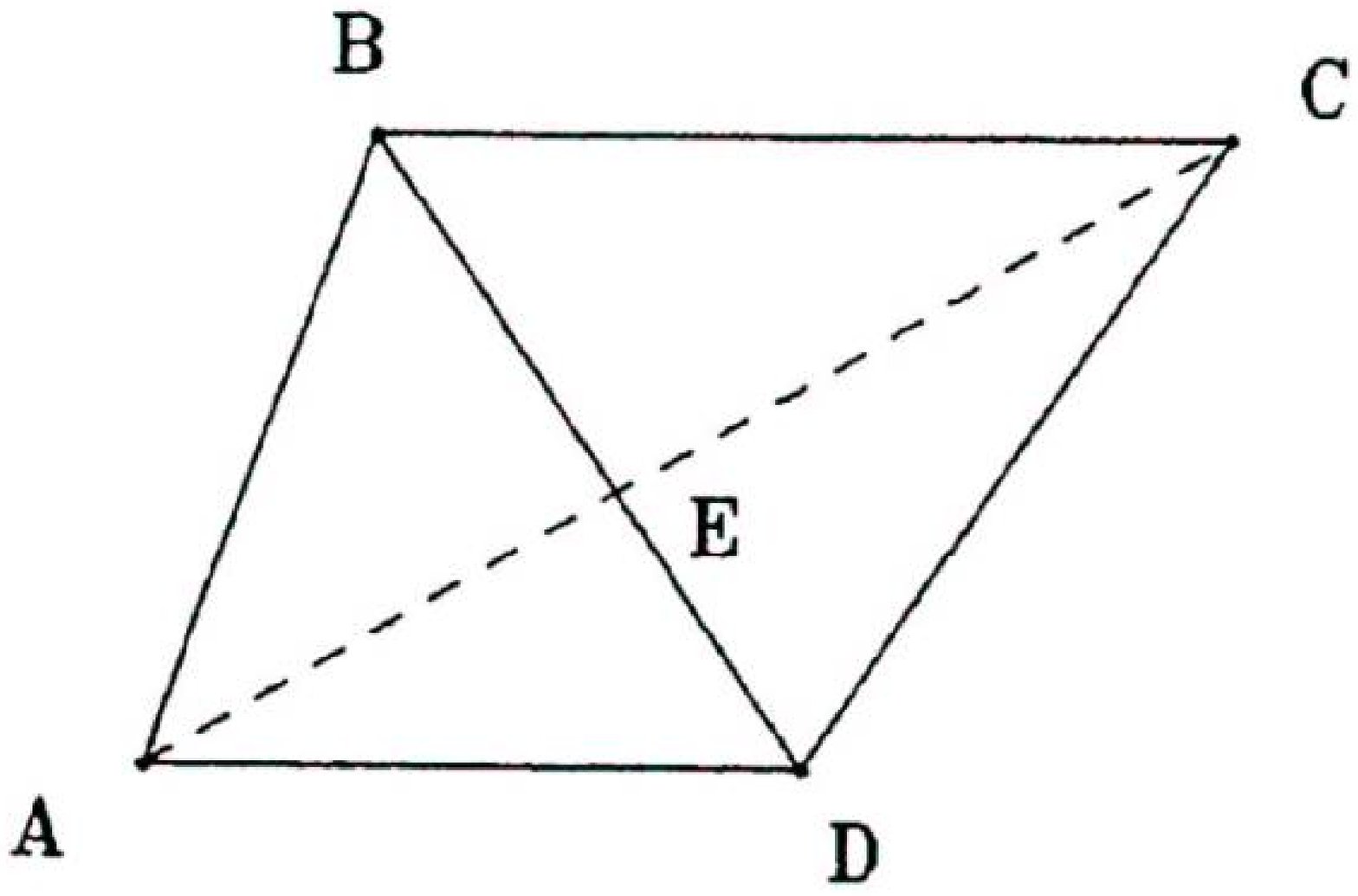,width=3in}}
\centerline{\psfig{figure=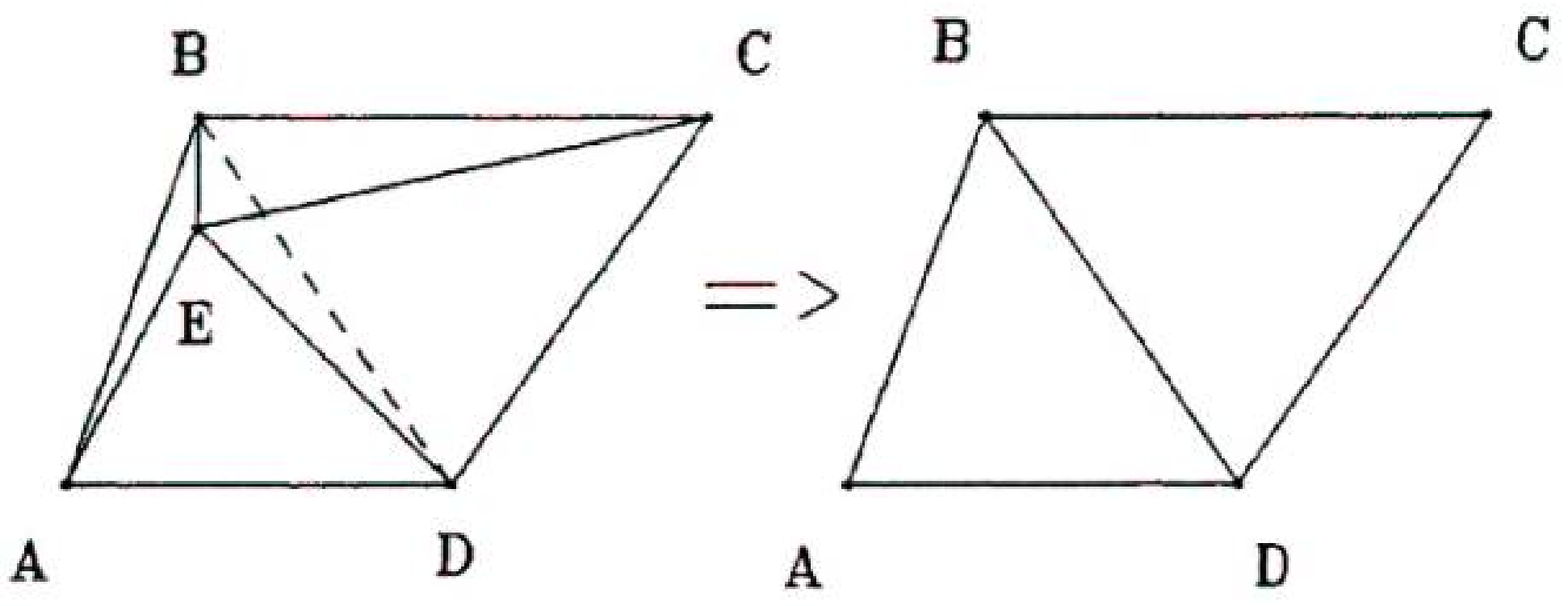,width=6in}}
\caption
{{\small
Elementary reconstructions: (Left up) BD connection is introduced
instead of AC connection. The total number of knots
and cells in the grid is not changed.
(Right up) Adding a knot at the middle of the connection:
the knot E is added to the existing knots ABCD
in the middle of BD connection, 2 connections
AE and EC appear, and the total number of cells is
increased by 2 cells.
(Down) Removing a knot: the knot E is removed from the grid,
and the total number of cells is decreased by 2 cells, from \cite{ard96}
}}
\label{rec}
\end{figure}

\noindent
The scheme is implicit. It was developed, and its stability and convergence
was investigated by the group of N.V.Ardeljan in Moscow State University.
The scheme is fully conservative, providing
conservation of the mass, momentum and total energy, and giving
correct calculation of the transitions between different types of energies.
Matrices for the transformation are symmetrical.
The difference scheme is Lagrangian, with triangular grid, and possibility of grid
reconstruction, see Fig.\ref{rec}. The angular momentum conserves automatically in this scheme.

\begin{figure}
\centerline{\psfig{figure=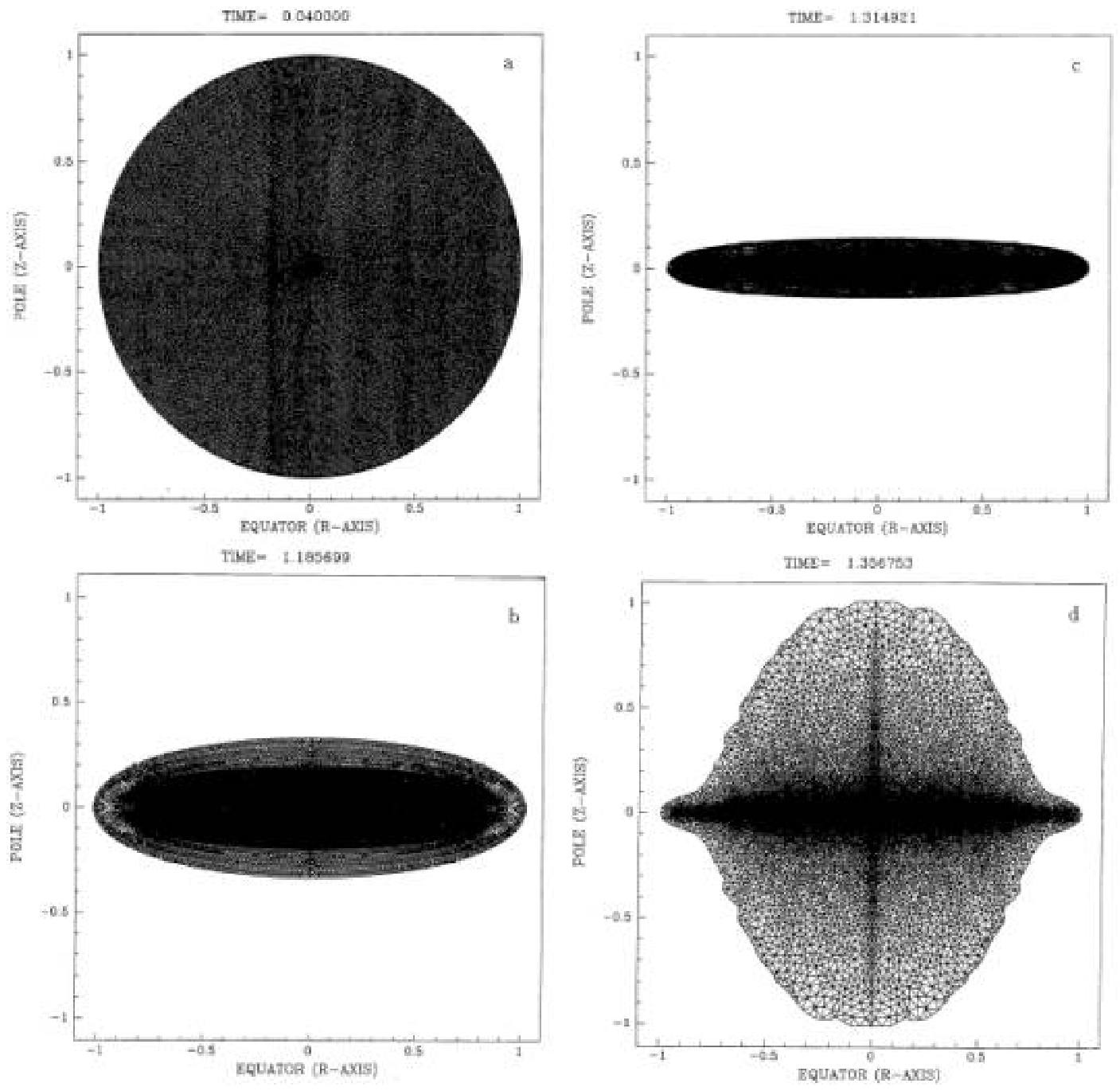,width=6in}}
\centerline{\psfig{figure=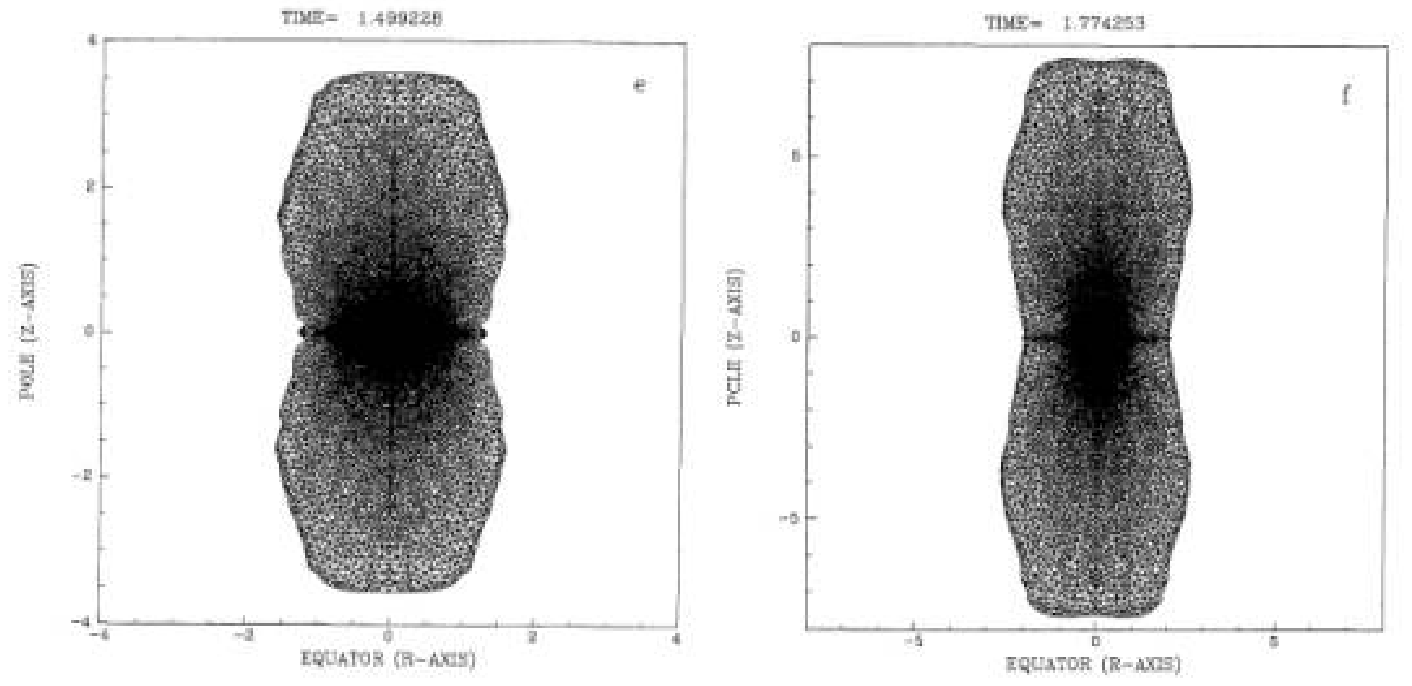,width=6in}}
\caption
{{\small
Time evolution of the shape of the rotating collapsing gas cloud, from \cite{ard96}.
}}
\label{uncloud}
\end{figure}

The astrophysical application of this scheme was started by
calculations of a collapse of a non-magnetized rotating
protostellar cloud, important in the problem of star formation
\cite{ard96}.
 Rigidly rotating  uniform gas sphere was considered  for
initial conditions.
The cloud, as a rigidly rotating
uniform gas sphere, was taken with the following initial parameters:

\begin{equation}
\label{eq1}
\rho=1.492\cdot 10^{-17}g/cm^3,\quad
p=1.548\cdot 10^{-10}dyn/cm^2,\quad
r=3.81\cdot10^{16}cm,
\end{equation}
$$\omega=2.008\cdot10^{-12}rad/sec, \quad
M=1.73M_\odot=3.457\cdot10^{33}g,\quad
\gamma={5/3},\quad u^r=u^z=0 .$$
On the outer boundary the pressure was taken to be equal  to a small constant
 $(p=0.87 \cdot 10^{-13}$ dyn/cm$^2$).
At the outer boundary of the cloud, the gravitational potential $\Phi$ is
defined by the integral Poisson formula, using the expression for the volume
potential. The results of calculations, showing formation of the bounce shock,
and subsequent outburst are represented in Fig.\ref{uncloud} from \cite{ard96}.

\section{Presupernova Core Collapse}
Collapse and formation of a rapidly and differentially rotating neutron star
had been calculated in \cite{ard04} in non-magnetized approximation.
Equations of state takes into account degeneracy of electrons and neutrons,
relativity for the electrons, nuclear transitions and nuclear interactions.
Temperature effects were taken into account approximately by the addition
of  the pressure of  radiation and of an ideal gas.
Neutrino losses and iron dissociation were taken into account in the energy equations.
A cool white dwarf was considered at the stability limit with a mass
equal to the Chandrasekhar limit.
To obtain the collapse we increase the density at each point by 20\%
and  switch on a uniform rotation.

The hydrodynamical equations with gravity for modelling the
nonstationary processes in rotating gaseous bodies are:

\begin{equation}
\frac{{\rm d} {\bf x}} {{\rm d} t} = {\bf v}, \quad
\frac{{\rm d} \rho} {{\rm d} t} +
\rho \nabla \cdot {\bf v} = 0,  \quad
\rho \frac{{\rm d} {\bf v}}{{\rm d} t} =-\nabla
\left(P \right)  - \rho  \nabla \Phi,  \label{eq2}
\end{equation}
$$
\rho \frac{{\rm d} \varepsilon}{{\rm d} t} +P \nabla \cdot {\bf v}+\rho F(\rho,T)=0,\quad
\Delta \Phi=4 \pi G \rho.
$$
Here the equation of state and the function of neutrino losses have been
represented by following functions

\begin{equation}
  P \equiv P(\rho,T)=P_0(\rho)+\rho \Re T + \frac {\sigma T^4} {3},\quad
  \varepsilon=\varepsilon_0(\rho)+\frac{3}{2}
\Re T +\frac{\sigma T^4}{\rho}+\varepsilon_{Fe}(\rho,T),
\label{eq3}
\end{equation}
where $\varepsilon_{Fe}(\rho,T)$ is the iron dissociation energy, F($\rho$,T)
is the function of neutrino losses, among which only the URCA process
losses are important. Other losses, such as pair annihilation,
photo production of neutrino, plasma neutrino were also included
in the calculations.

\begin{equation}\label{pressure}
P_0(\rho)=\left\{
\begin{array}{rclcccccc}
P_0^{(1)}&=&b_1\rho^{1/3}/(1+c_1\rho^{1/3}),& &&\rho&\leq&\rho_1,&\\
P_0^{(k)}&=&a\cdot 10^{b_k({\textrm{lg}}\rho-8.419)^{c_k}}&
\rho_{k-1}&\leq&\rho&\leq&\rho_k,&k=\overline{2,6}\\
\end{array}
\right.
\end{equation}
\begin{equation}\label{inten}
\varepsilon=\varepsilon_0(\rho)+\frac{3}{2}
\Re T +\frac{\sigma T^4}{\rho}+\varepsilon_{Fe}(\rho,T), \quad
\varepsilon_0(\rho)=\int\limits_0^\rho \frac{P_0(\tilde{\rho})}
{\tilde{\rho}^2}\textrm{d}\tilde{\rho},\quad
\varepsilon_{Fe}(\rho,T)=\frac{E_{b,Fe}}{A\>m_p}
  \left(\frac{T-T_{0Fe}}{T_{1Fe}-T_{0Fe}}\right).
\end{equation}

\begin{equation}\label{urca}
  f(\rho,T)=\frac {1.3 \cdot 10^9 {\textrm {\ae}}(\overline{T})\overline{T}^6}
  {1+(7.1\cdot 10^{-5}\rho /\overline{T}^3)^{{2}/{5}}}\quad
  {\textrm {erg}} \cdot {\textrm{g}}^{-1} \cdot {\textrm {s}}^{-1},\quad
  \overline {T}=T\cdot 10^{-9},
\end{equation}
\begin{equation}
{\textrm {\ae (T)}}=\left\{
\begin{array}{rclcccccc}
1,&\overline{T}<7,\\
664.31+51.024 (\overline{T}-20), & 7\leq \overline{T} \leq 20,\\
664.31, & \overline{T}>20,
\end{array}
\right.
\end{equation}
The URCA losses $f(\rho, T)$ have been described by formula \cite{bkps76},
obtained by approximation of tables from
\cite{iin69}. Other losses (unimportant) are included in $Q_{tot}$.

\begin{equation}\label{neuttot}
  F(\rho,T)=(f(\rho,T)+Q_{tot})e^{-\frac{\tau_\nu}{10}}.
\end{equation}

\begin{figure}
\centerline{\psfig{figure=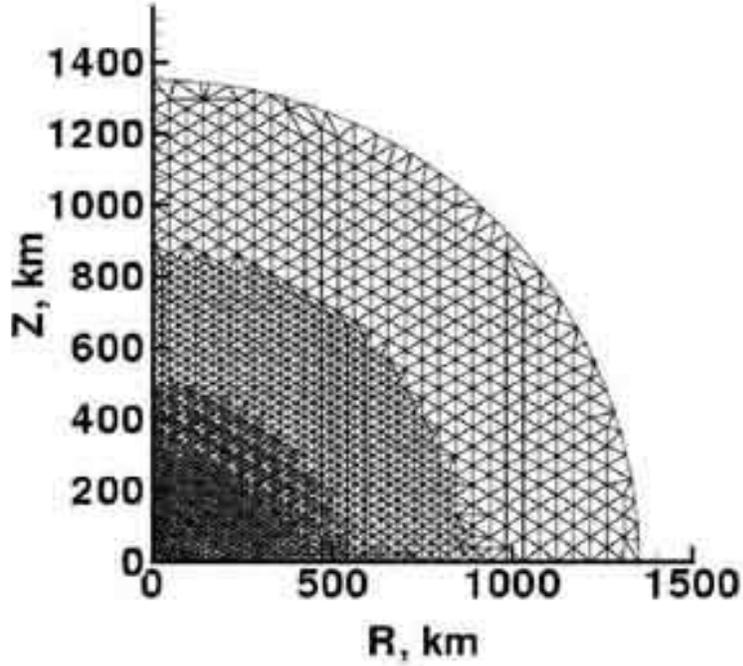,width=4.2in}}
\caption
{{\small
The triangle grid at the beginning, $t=0$, from \cite{ard04}.
}}
\label{grid}
\end{figure}

Neutrino absorption is important in deeper layers of the neutron star.
It is taken into account implicitly, by multiplying the transparent
neutrino flux by the multiplier $e^{-\frac{\tau_\nu}{10}}$, where the
effective neutrino optical depth $\tau_{\nu}$ is calculated using
local density gradients \cite{ard04},\cite{ard05}.
Initial state was chosen as a
  spherically symmetric star, with a mass
  20\% larger than the limiting mass of the corresponding white dwarf
   $M=1,0042M_\odot$ + 20\%,
and rotating uniformly, with an angular velocity 2.519 (1/sec).
The temperature distribution was taken in the form
  $T=\delta\rho^{2/3}$.
The initial triangular grid covering a quarter of the circle, where equations are solved,
is represented in Fig.\ref{grid}.

\begin{figure}
\centerline{\psfig{figure=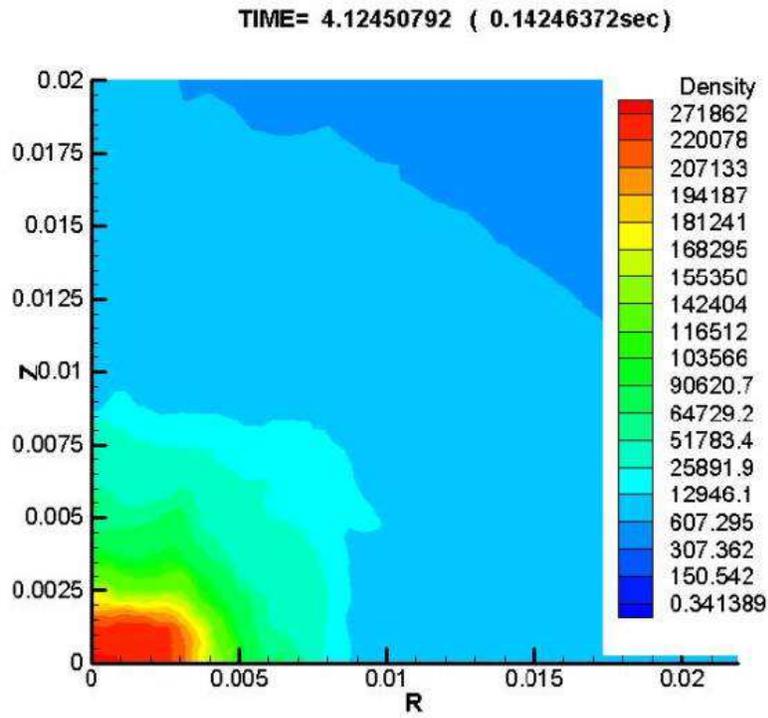,width=4.5in}}
\caption
{{\small
Density distribution in the central region
at maximum compression state $t=0.1425$ sec, from \cite{ard04}.
}}
\label{dense}
\end{figure}

\begin{figure}
\centerline{\psfig{figure=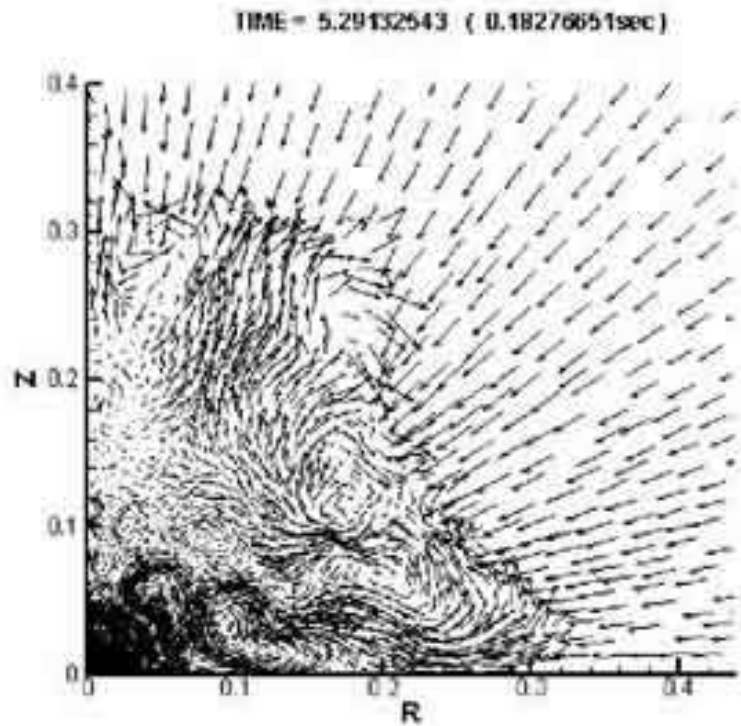,width=4.5in}}
\caption
{{\small
Velocity field, showing mixing of matter behind
the shock at $t=0.1828$, from \cite{ard04}.
}}
\label{mix}
\end{figure}

\begin{figure}
\centerline{\psfig{figure=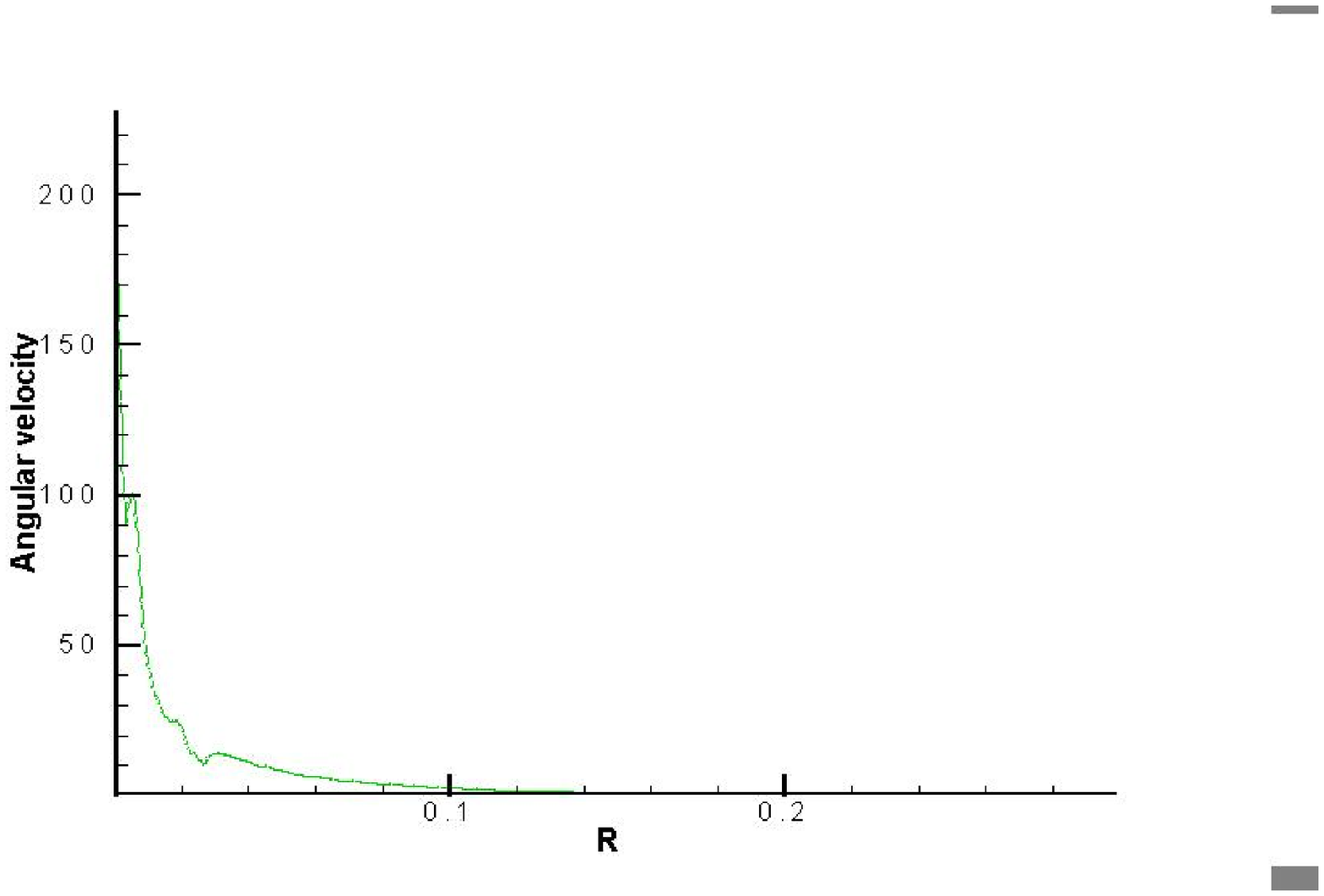,width=6in}}
\caption
{{\small
Angular velocity distribution in the equatorial plane
of the new born neutron star at $t=0.261$ after the beginning
of the collapse.
}}
\label{ang}
\end{figure}

Results of calculations are represented in Figs.\ref{dense} - \ref{ang}.
The collapse is stopped due to formation of a stable neutron core,
rotating almost uniformly. The shock is reflected from the core,
leaving behind a hot plasma, where vortexes are generated.
It leads to a strong mixing, what is important for the interpretation of
observations of the light curve of SN 1987A. The light curve, corresponding
to the radioactive decay of
$^{56}$Ni $\rightarrow$ $^{56}$Co $\rightarrow$ $^{56}$Fe was observed in this
object, what could be explained by enrichment of the outbursting matter
due to mixing with deeper layers.

The main result of these calculations
is a self-consistent model of the rapidly rotating neutron star
with a differentially rotating envelope. The adjusting parameters
are chosen to reproduce the calculations with a refined account of the
neutrino transport \cite{jan03}. In correspondence with these calculations
we have obtained that the bounce shock wave and neutrino deposition
do not produce SN explosion.
Distribution of the angular velocity is represented in Fig.\ref{ang}.
The period of rotation of the almost uniformly rotating core of the young
neutron star is about  0.001 sec.

\section{2-D magnetorotational supernova}

\begin{figure}
\centerline{\psfig{figure=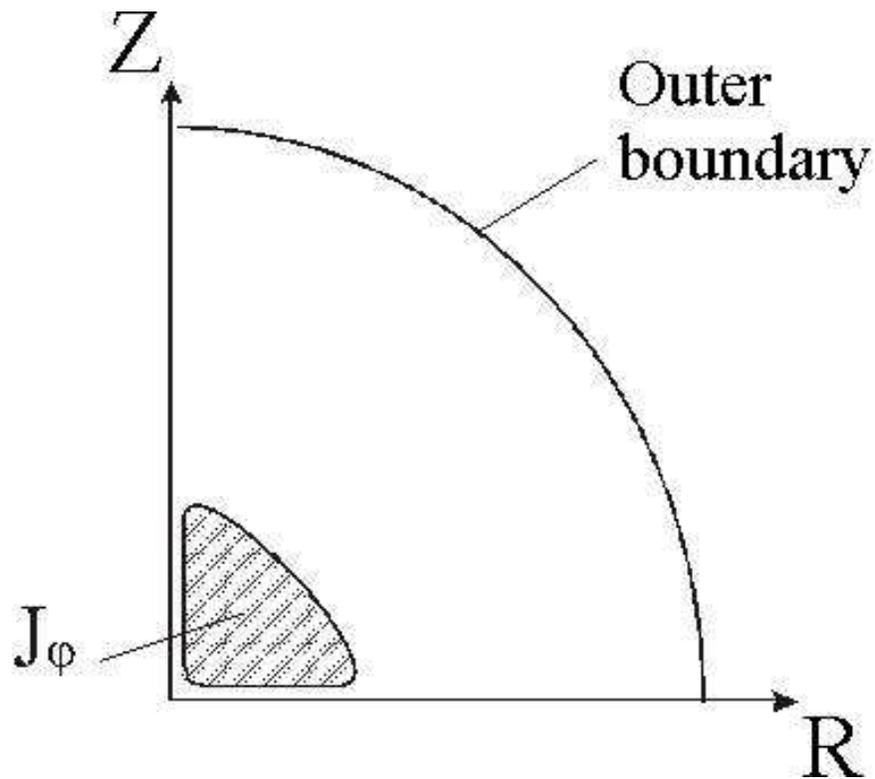,width=4.5in}}
\caption
{{\small
Initial, at $t=0$, current distribution in the upper hemisphere.
Dipole-like field is obtained when currents in both hemispheres
have the same sign, and quadrupole-like field appears when currents
have opposite signs.}}
\label{current}
\end{figure}

\begin{figure}
\centerline{\psfig{figure=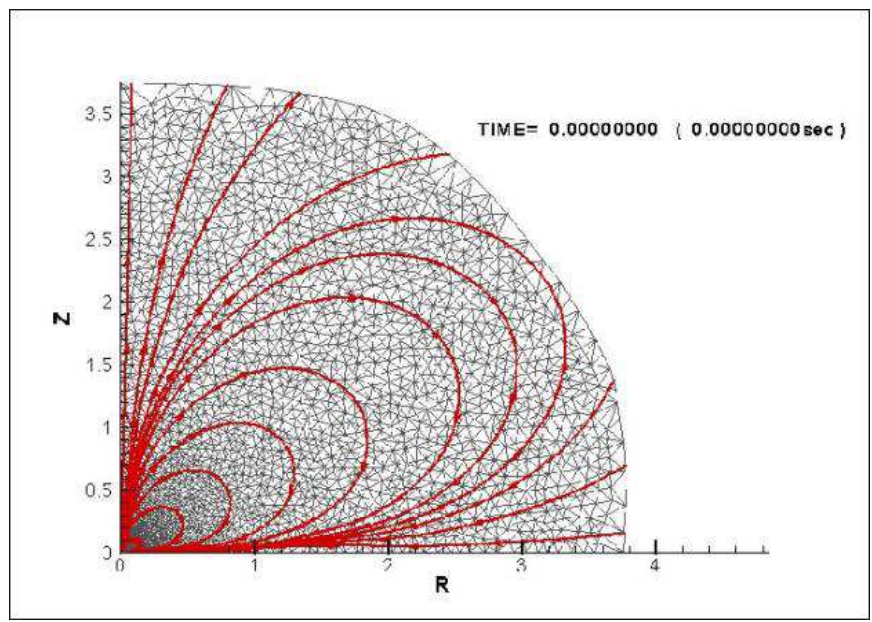,width=4.5in}}
\caption
{{\small
Initial quadrupole-like magnetic field, at $t=0$.
}}
\label{quadr}
\end{figure}

\begin{figure}
\centerline{\psfig{figure=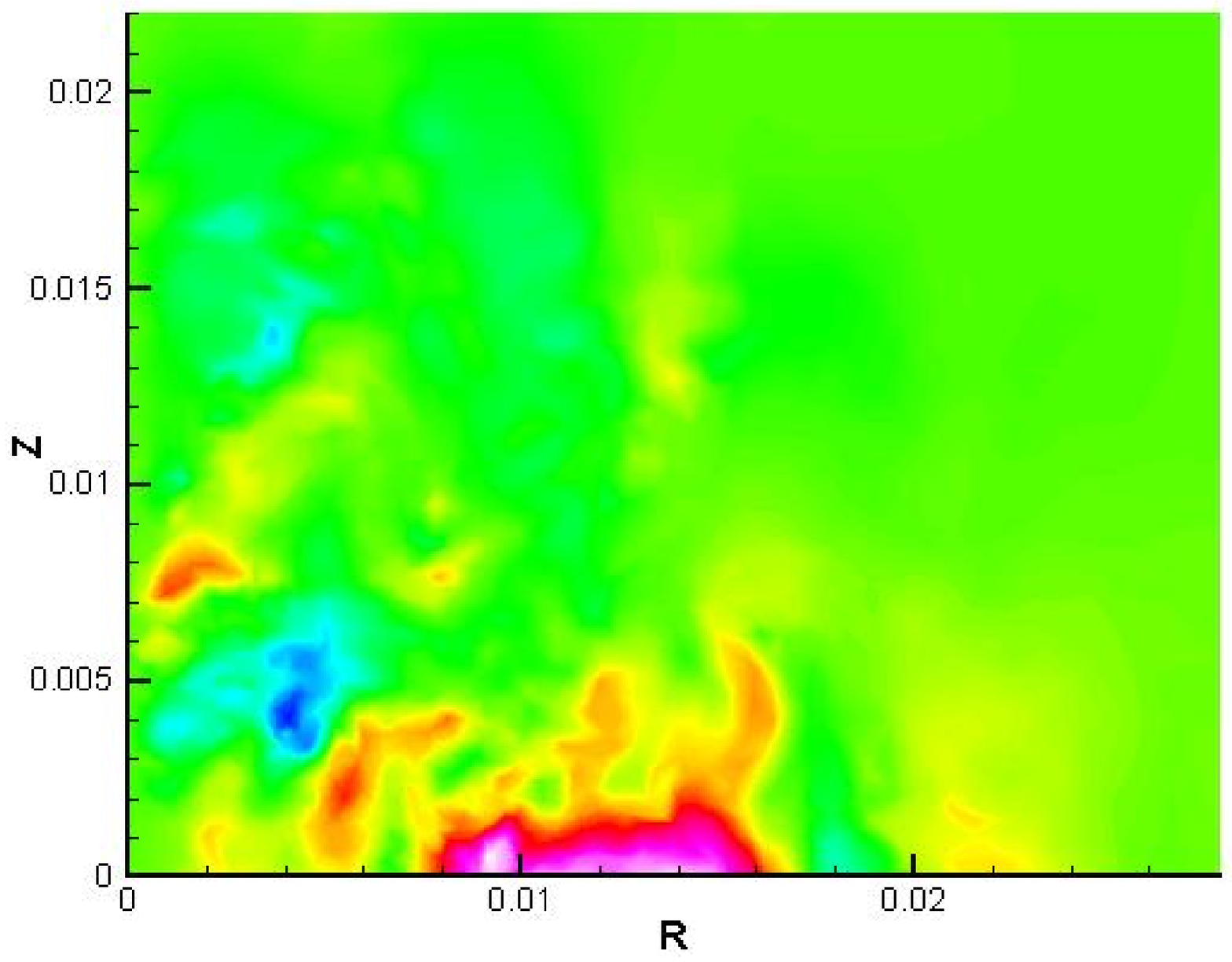,width=4.5in}}
\caption
{{\small
Toroidal magnetic field distribution at the moment of
its maximal energy.
}}
\label{tormax}
\end{figure}

\begin{figure}
\centerline{\psfig{figure=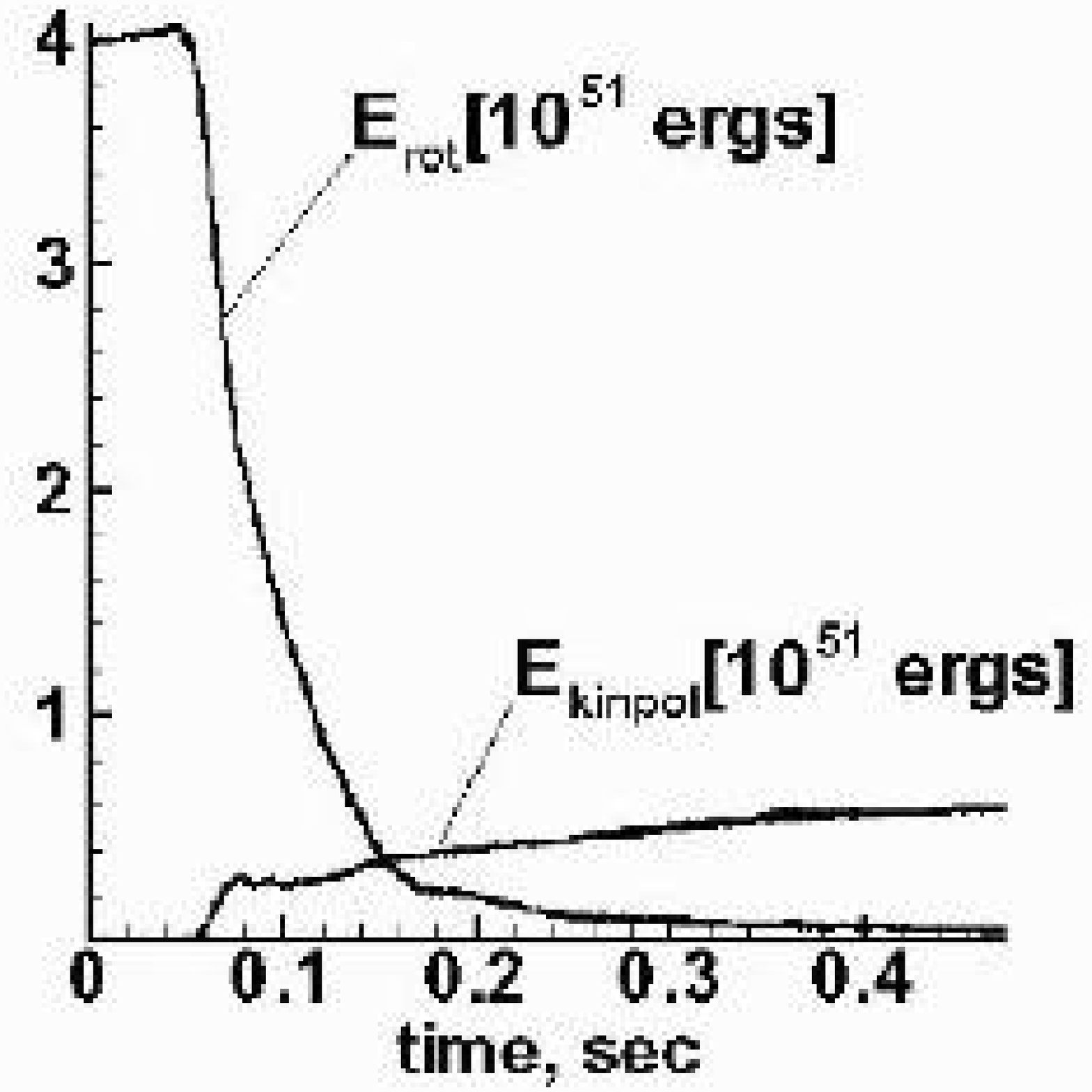,width=3in}
            \psfig{figure=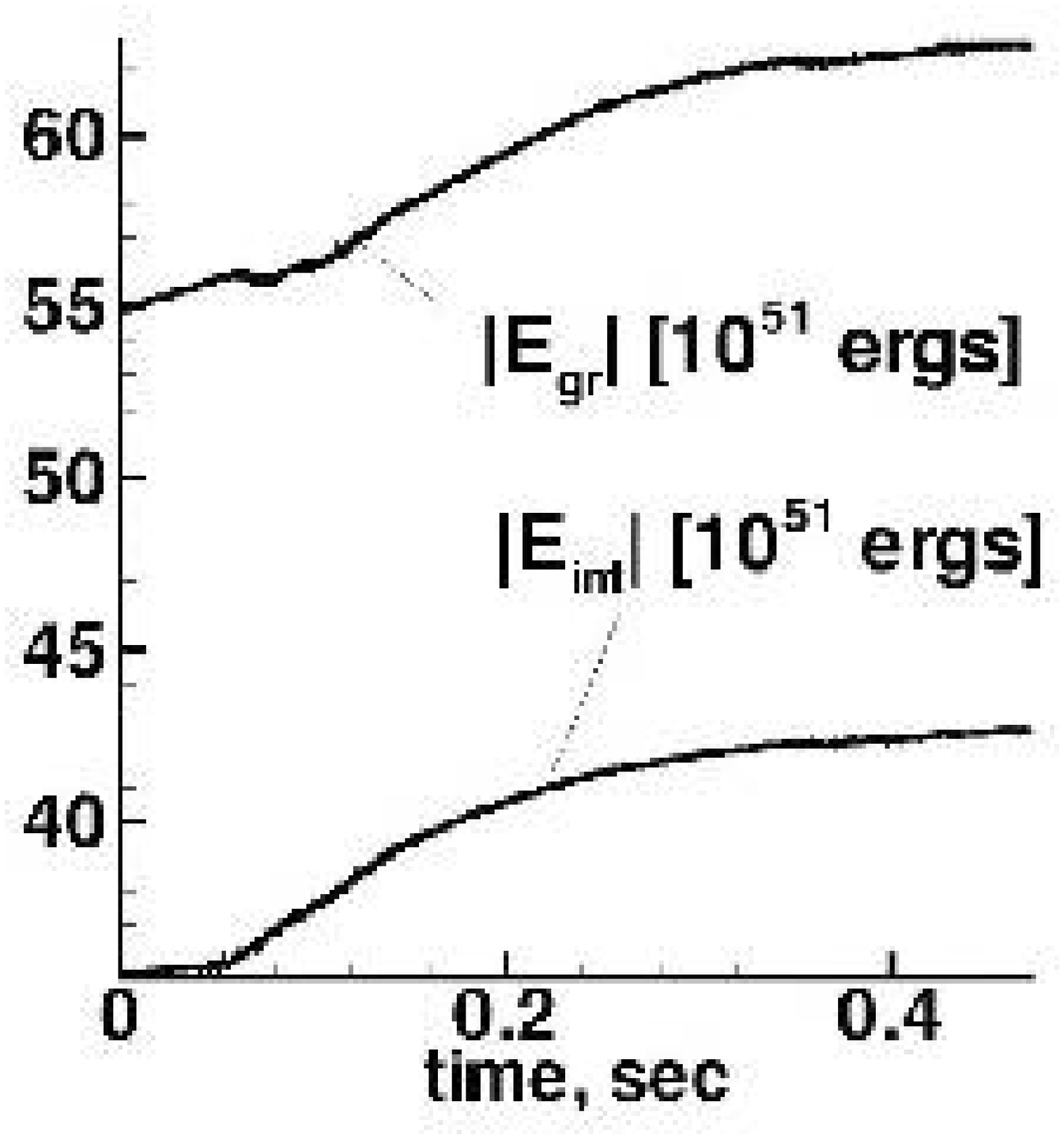,width=3in}}
\caption
{{\small
  Time dependence of the gravitational, internal, rotational, and kinetic poloidal
energies
 of the star during magnetorotational explosion with a quadrupole-like field,
 from \cite{ard05}.}}
\label{engravetal}
\end{figure}

\begin{figure}
\centerline{\psfig{figure=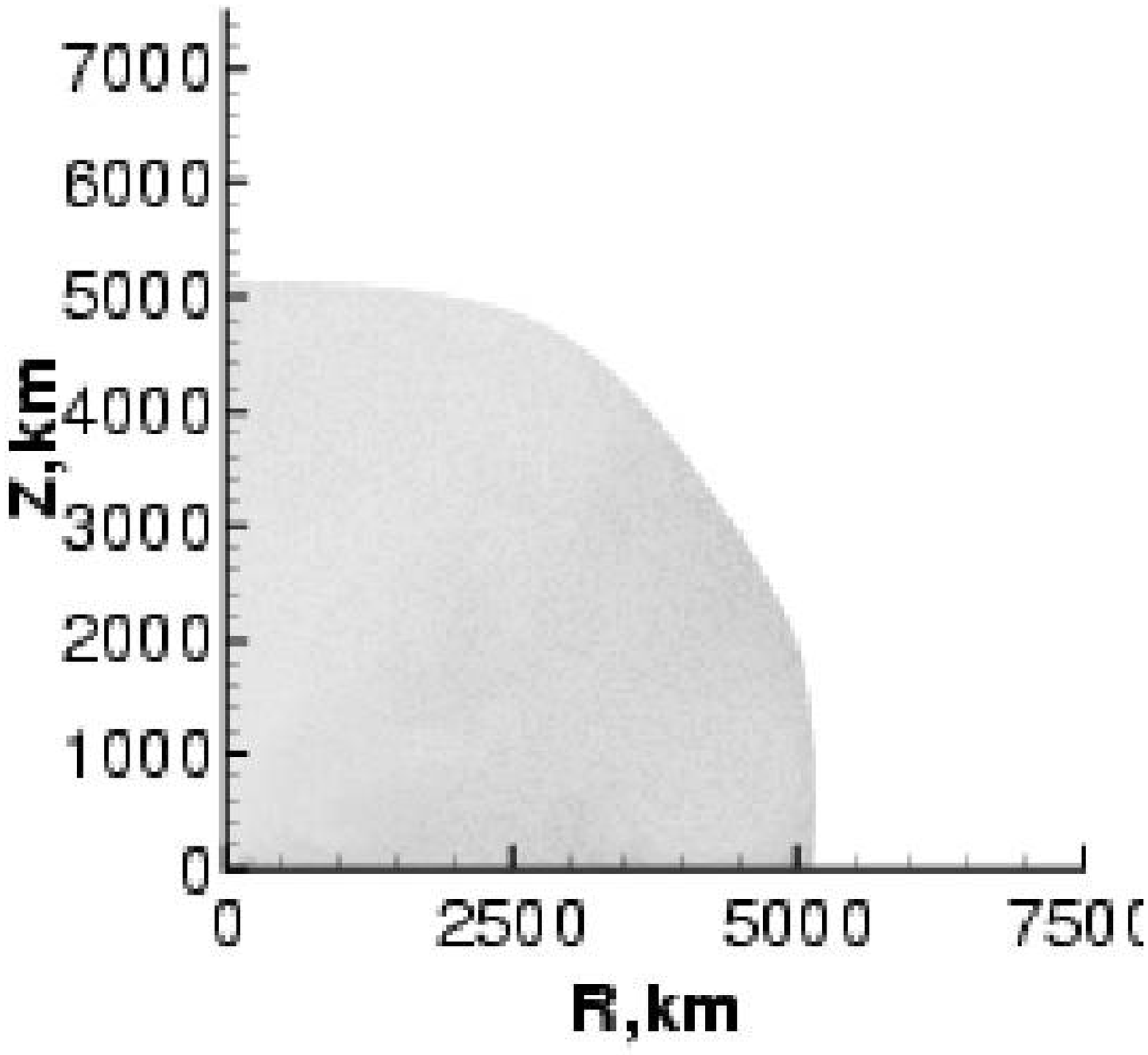,width=3in}
              \psfig{figure=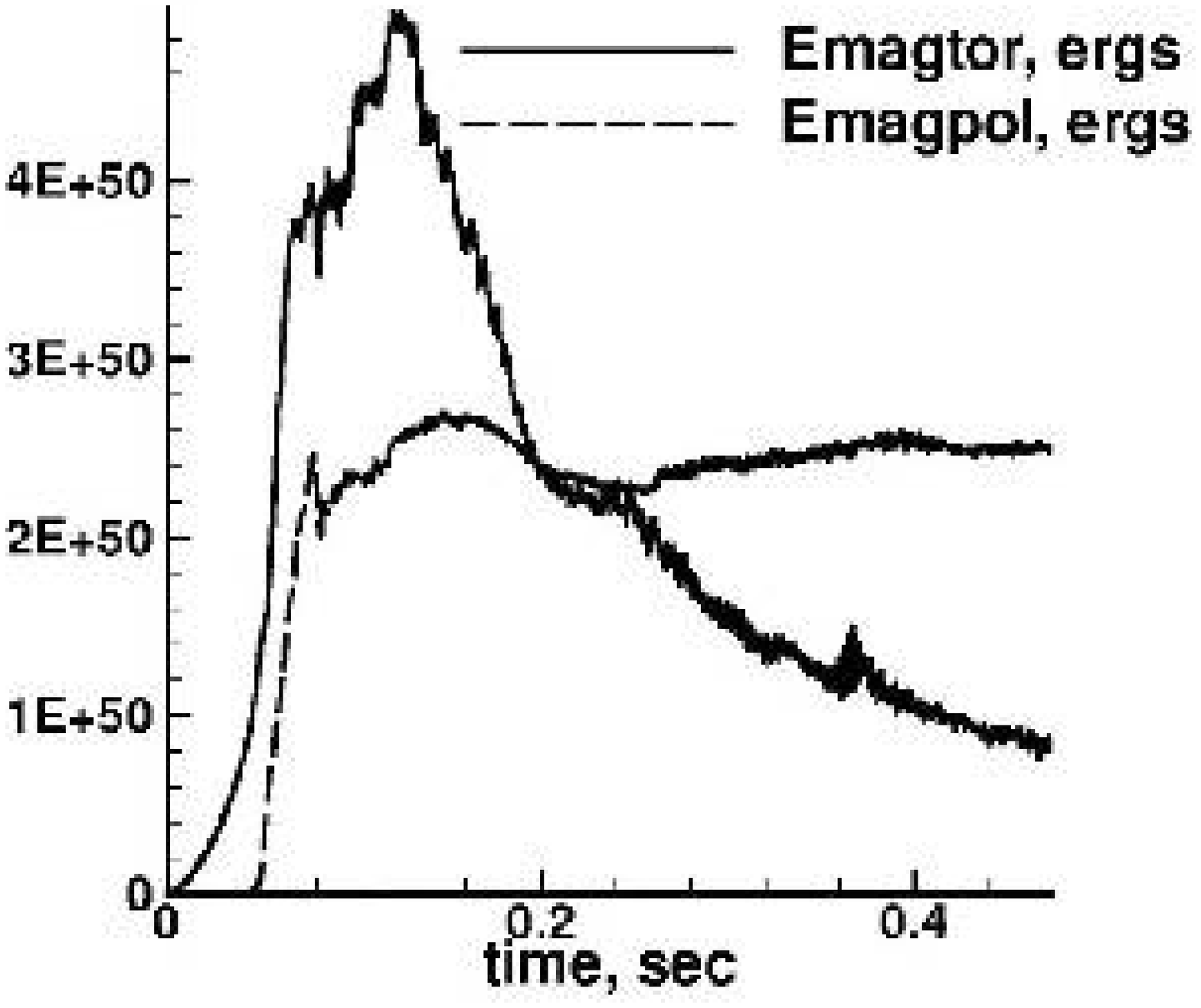,width=3in}}
\caption
{{\small
{\bf Right} Time dependencies of the magnetic
(toroidal and poloidal separately) energies
of the star during magnetorotational explosion with a quadrupole-like field.
{\bf Left} The specific angular momentum $v_\varphi r$ distribution
for the time moment $t=0.07s$, from \cite{ard05}}}
\label{enmag}
\end{figure}

\begin{figure}
\centerline{\psfig{figure=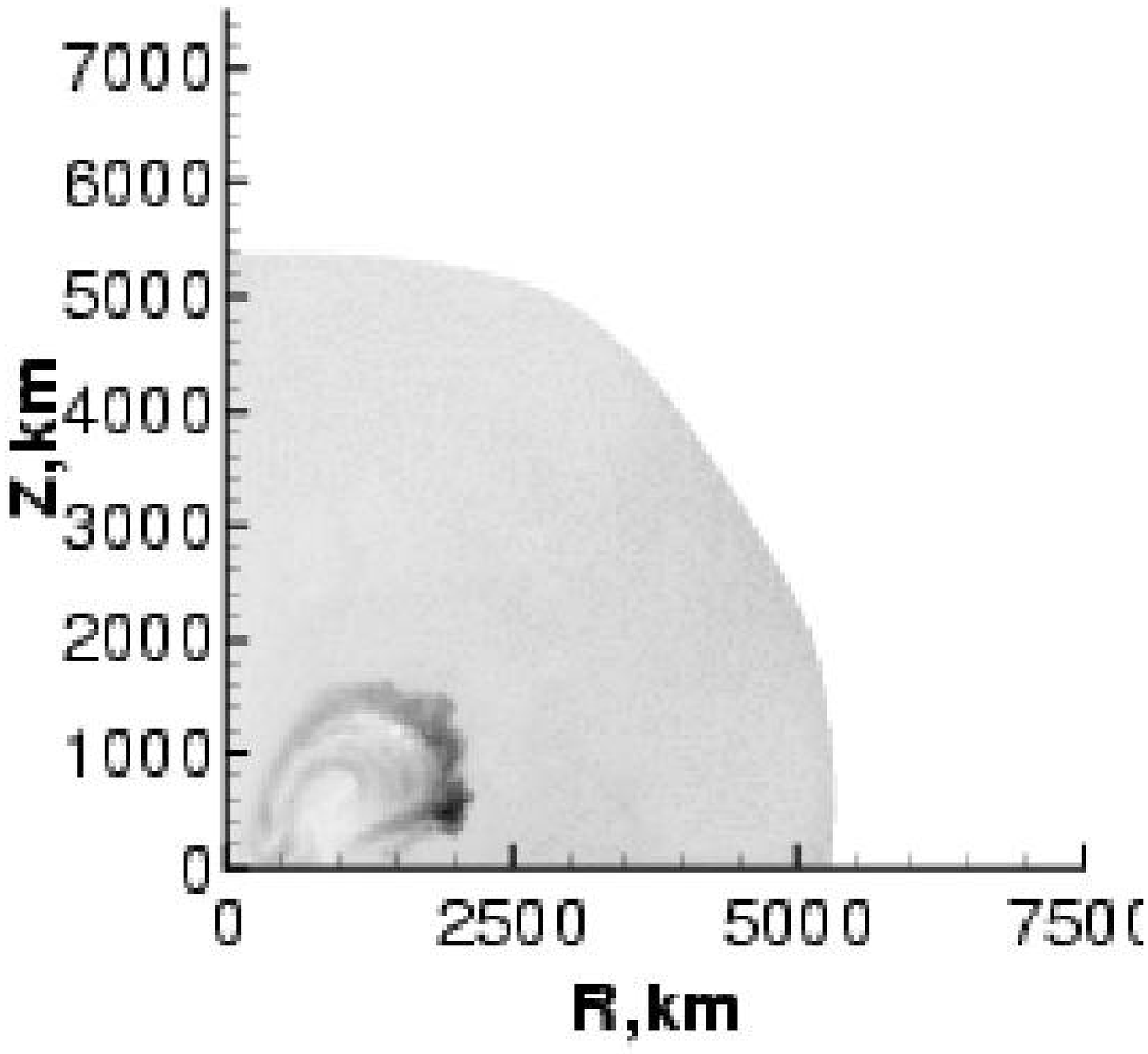,width=3in}
             \psfig{figure=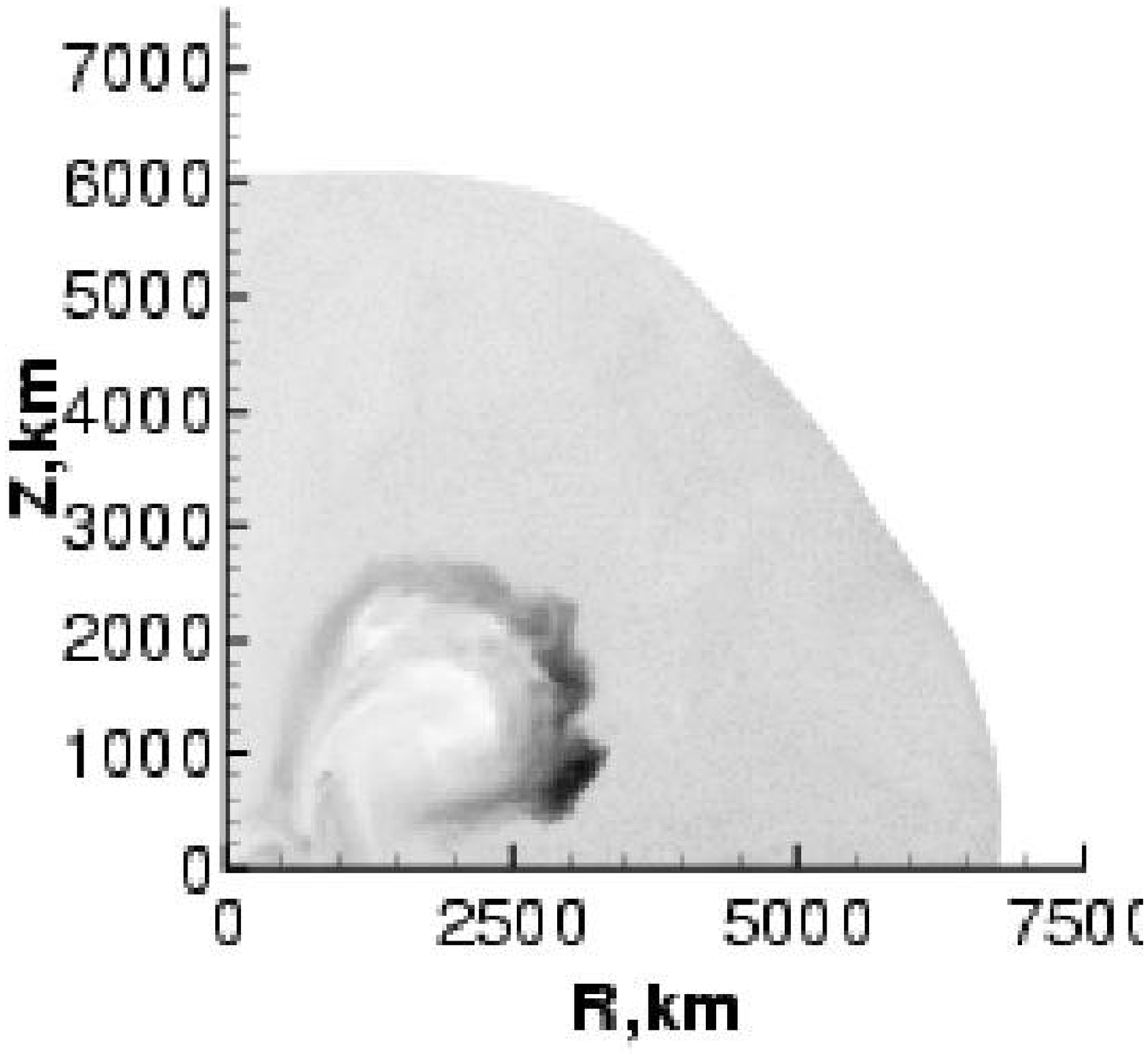,width=3in}}
\caption
{{\small
Time evolution of the specific angular momentum $v_\varphi r$
for the time moments $t=0.20s,\> 0.30s$.
  The darker parts of the plots correspond to the
  higher specific angular momentum, from \cite{ard05}.}}
\label{angmom}
\end{figure}

Calculations of magnetorotational core-collapse supernova have been
performed in \cite{ard05}. Magnetohydrodynamic (MHD) equations with
self-gravitation, and infinite conductivity have been solved using the same
numerical scheme as described above. The set of equations is the following:

\begin{eqnarray}
\frac{{\rm d} {\bf x}} {{\rm d} t} = {\bf v}, \quad
\frac{{\rm d} \rho} {{\rm d} t} +
\rho \nabla \cdot {\bf v} = 0,  \quad
\rho \frac{{\rm d} {\bf v}}{{\rm d} t} =-{\rm grad}
\left(P+\frac{{\bf H} \cdot {\bf H}}{8\pi}\right) +
\frac {\nabla \cdot({\bf H} \otimes {\bf H})}{4\pi} -
\rho  \nabla \Phi, \nonumber\\
\rho \frac{{\rm d}}{{\rm d} t} \left(\frac{{\bf H}}{\rho}\right)
={\bf H} \cdot \nabla {\bf v},\>\,\,\,
\Delta \Phi=4 \pi G \rho,\quad
\rho \frac{{\rm d} \varepsilon}{{\rm d} t} +
P \nabla \cdot {\bf v}+\rho F(\rho,T)=0, \quad
P=P(\rho,T),\> \varepsilon=\varepsilon(\rho,T),
\label{magmain}
\end{eqnarray}
where $\frac {\rm d} {{\rm d} t} = \frac {\partial} {
\partial t} + {\bf v} \cdot \nabla$ is the total time
derivative, ${\bf x} = (r,\varphi , z)$, ${\bf v}=(v_r,v_\varphi,v_z)$ is the
 velocity
vector, $\rho$ is the density, $P$ is the pressure,  ${\bf
H}=(H_r,\> H_\varphi,\> H_z)$ is the magnetic field vector, $\Phi$ is
the gravitational potential, $\varepsilon$ is the internal energy, $G$ is
gravitational constant, ${\bf H} \otimes {\bf H}$ is the tensor
of rank 2, and
$F(\rho,T)$ is the rate of neutrino losses.
Equation of state and the function of neutrino losses have been the same as
in the previous section.
Additional condition here is ${\textrm div}{\bf H}=0$.
The problem has an axial symmetry ( $\frac{\partial}{\partial \phi}=0)$,
and the symmetry to the equatorial plane (z=0).
Boundary conditions are the following:

\begin{equation}
P=\rho=T=B_\phi=0\,\, \hbox{at the outer boundary};
 \quad v_r=j_r=B_r=0\,\, \hbox{at r = 0};
\end{equation}
$$
 v_\phi =j_\phi=B_\phi=0\,\,\hbox{at r = 0};\quad
  v_z=0,\quad
  \frac{\partial{B_z}}{\partial z}=0\,\,\hbox{(dipole-like)}
   \hbox{  or  } B_z=0 \,\,\hbox{(quadrupole-like)}\,\,\, \hbox{at z = 0}.
$$
Initial toroidal current $J_{\phi}$ (see Fig.\ref{current})
was taken at the initial moment (time started now from the stationary rotating
neutron star) producing $H_r, \,\, H_z$ according to Bio-Savara law
\begin{equation}
\label{biosavara}
{\bf H}=\frac 1 c \int\limits_V^{}
\frac {{\bf J} \times {\bf R}}{{R}^3}dV,
\end{equation}
Initial magnetic field of quadrupole-like symmetry is obtained
by opposite directions of the current in both hemispheres (Fig.\ref{quadr}).
Magnetic field is amplified due to twisting by the differential rotation,
and subsequent development of the magnetorotational instability.
The field distribution for initial quadrupole-like magnetic field with
$\alpha=10^{-6}$, at the moment of the maximal energy of the toroidal
magnetic field is represented in Fig.\ref{tormax}. Two dark areas: near the
equatorial plane, and around the axis at 45$^o$, show the regions with local
maxima of the toroidal magnetic field  $H_\phi^2$.
The maximal value of $H_\phi=2.5\cdot 10^{16}$ Gs was obtained in the
calculations. The magnetic field at the surface of the neutron star after the explosion is
$H=4 \cdot 10^{12}$ Gs.
\begin{figure}
\centerline{\psfig{figure=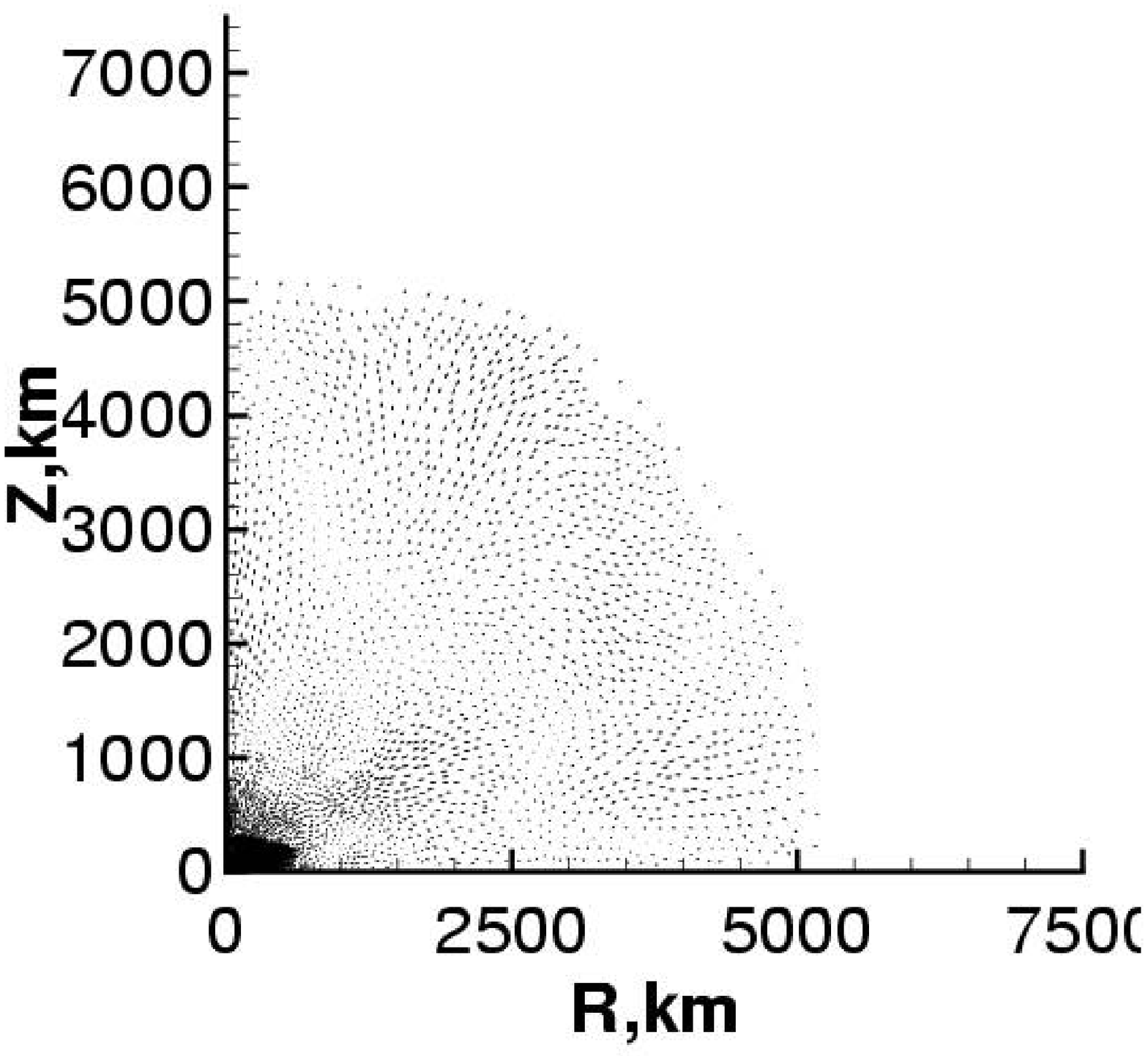,width=2.8in}
            \psfig{figure=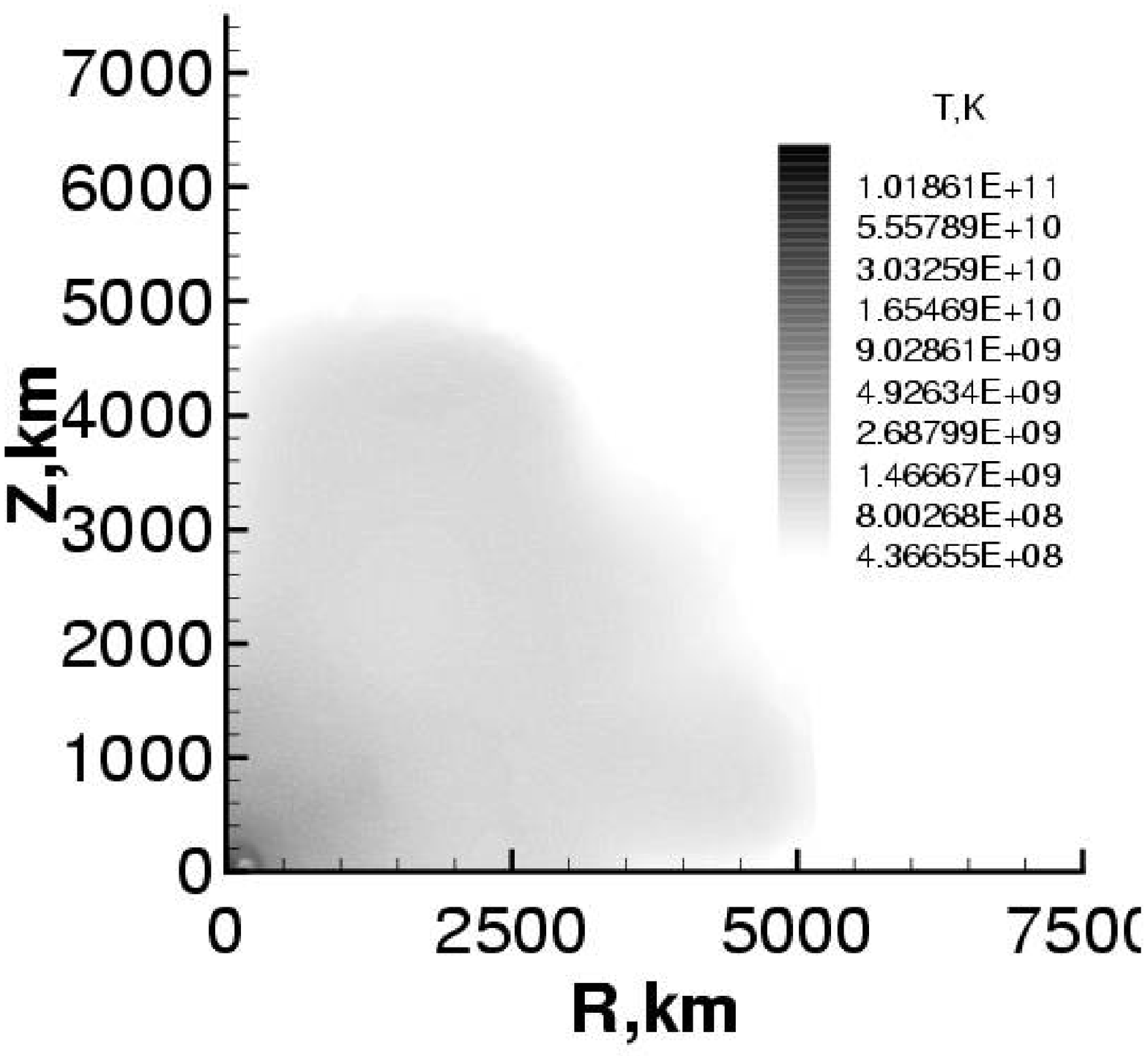,width=2.8in}}
\centerline{\psfig{figure=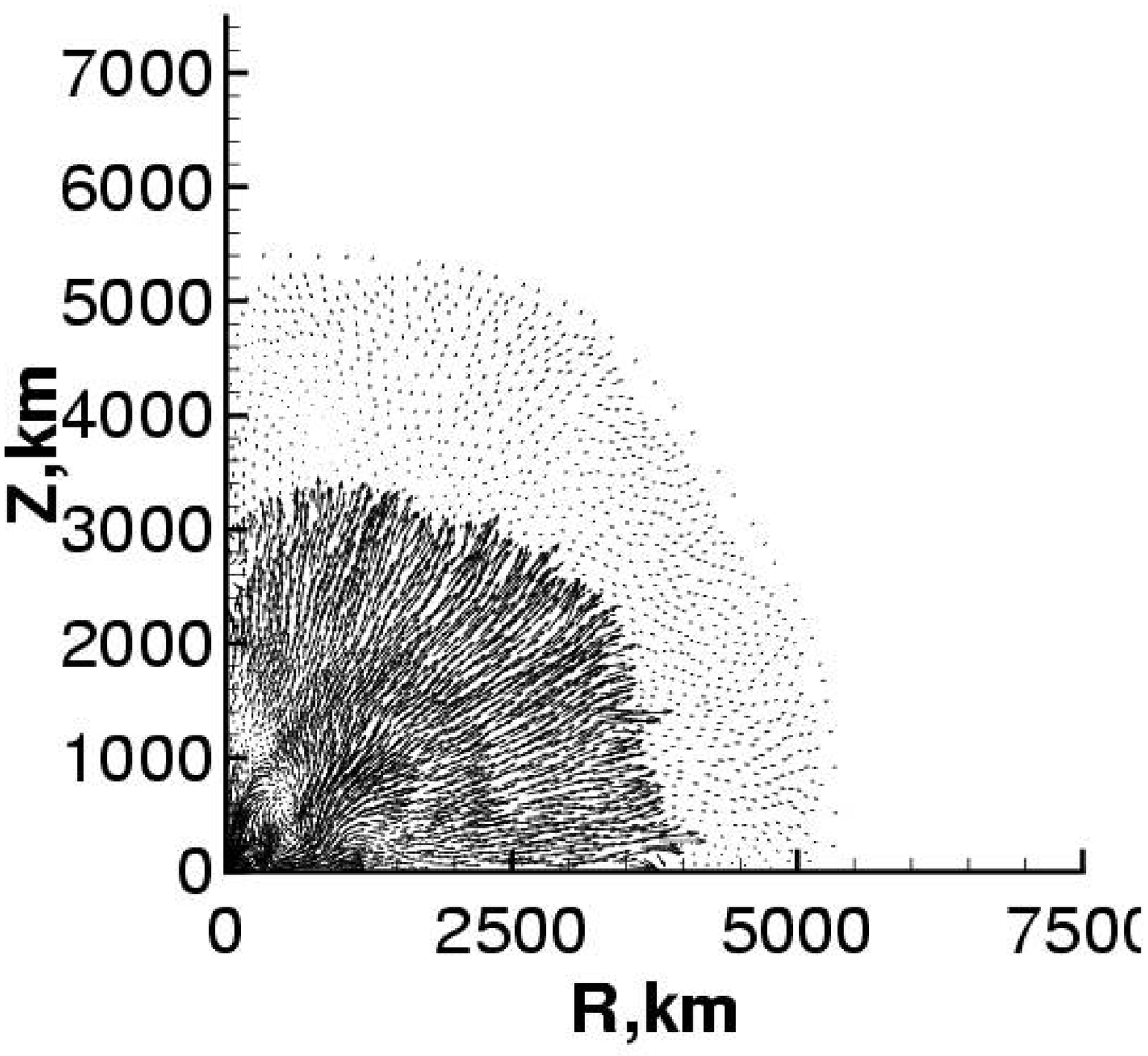,width=2.8in}
            \psfig{figure=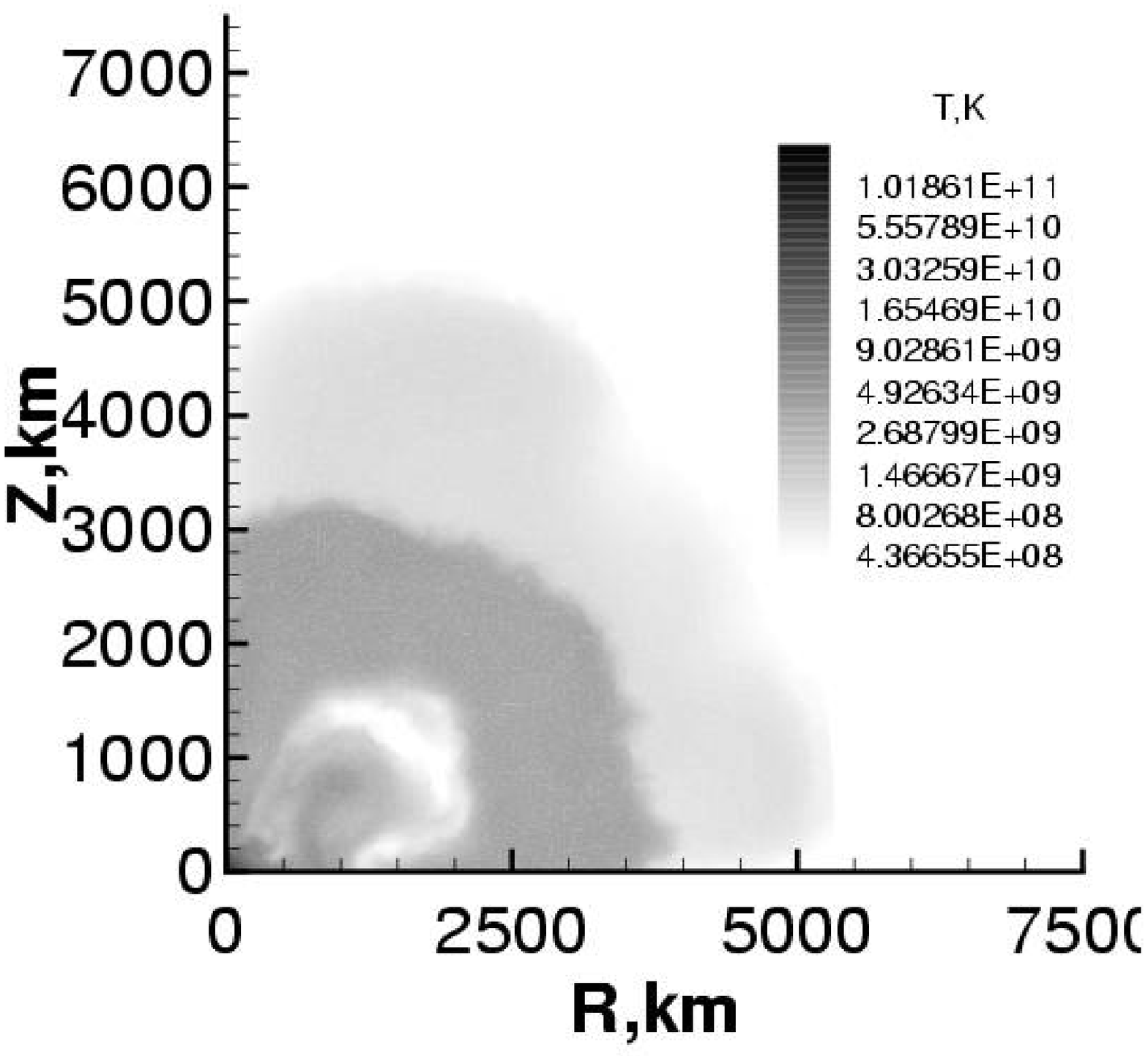,width=2.8in}}
\centerline{\psfig{figure=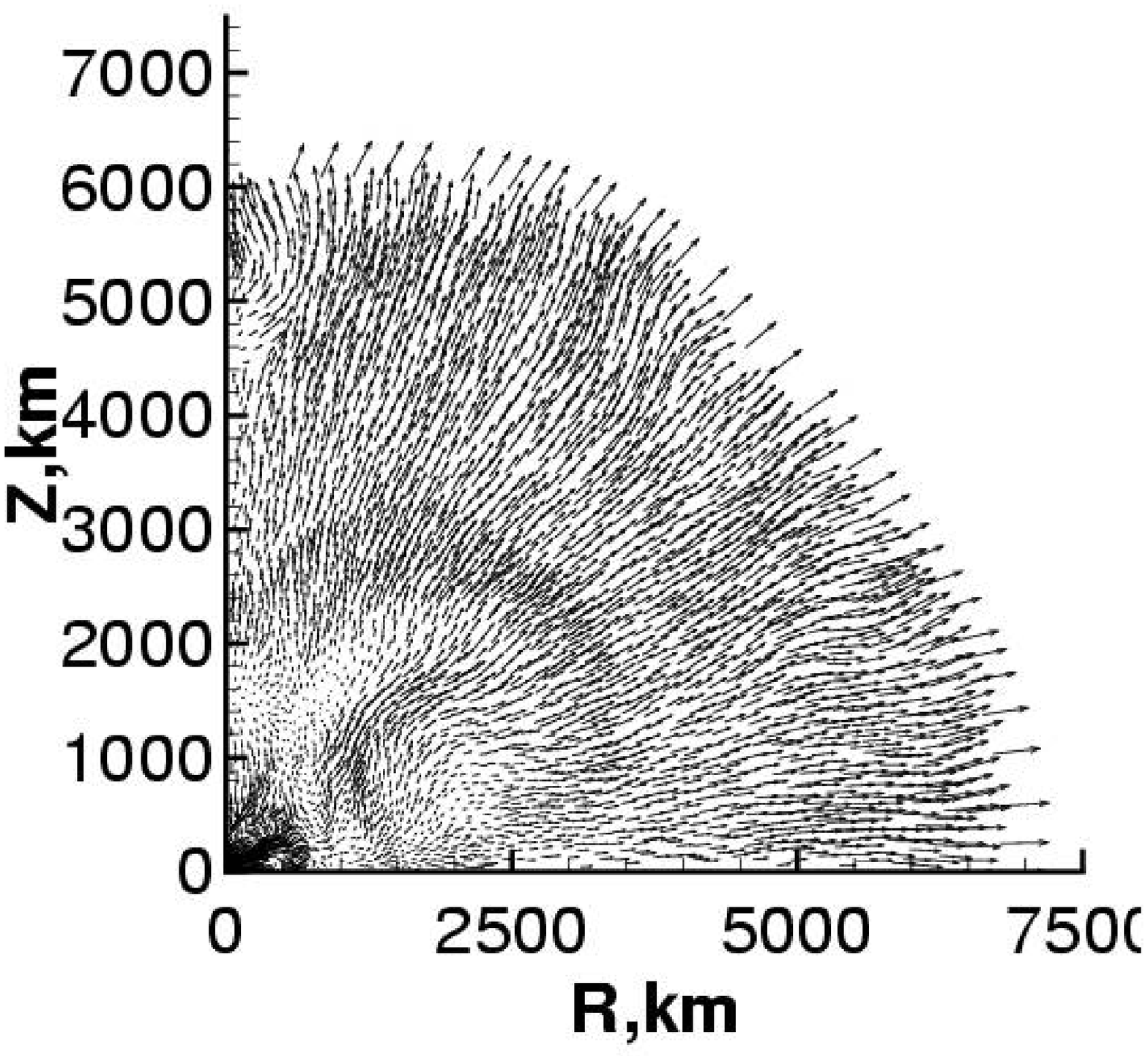,width=2.8in}
           \psfig{figure=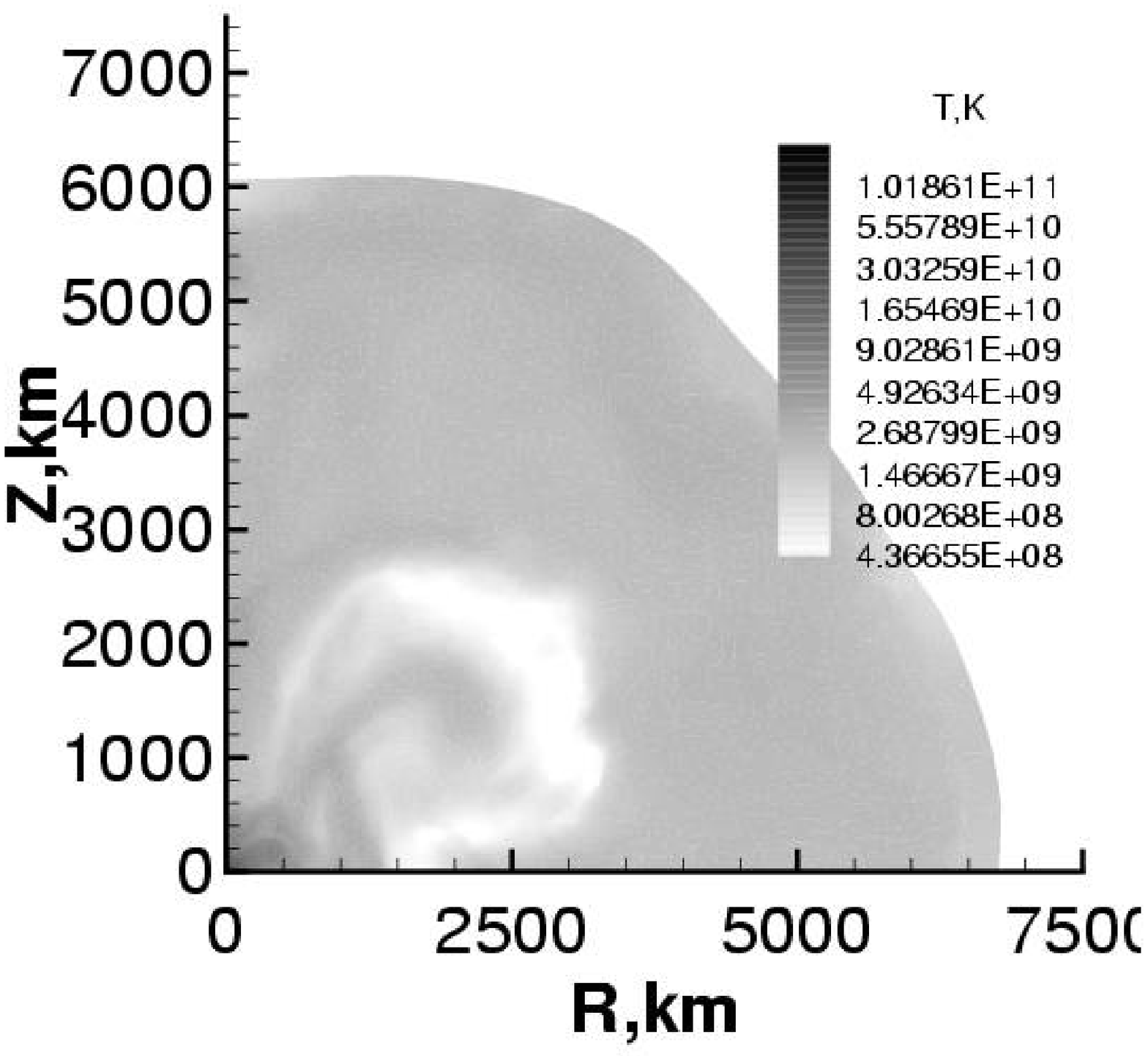,width=2.8in}}
\caption
{{\small
Time evolution of the velocity field (left column) and temperature
  (right column) for the time moments $t=0.07s,\>0.20s,\>0.30s$,
  from \cite{ard05}.}}
\label{tv}
\end{figure}
Temperature and velocity fields development with time is given in
Fig.\ref{tv}.
Time dependencies during the explosion of different types of the
energy: rotational energy, gravitational energy,
internal energy,
kinetic poloidal energy, are given in Figs.\ref{engravetal}.
The evolution of the magnetic poloidal energy and magnetic toroidal
 energies are given in the right side of Fig.\ref{enmag}. Distribution of the specific
 angular momentum $rv_\phi$ at different time moments is given in Fig.\ref{enmag} (left), and
 Fig.\ref{angmom}.

\begin{figure}
\centerline{\psfig{figure=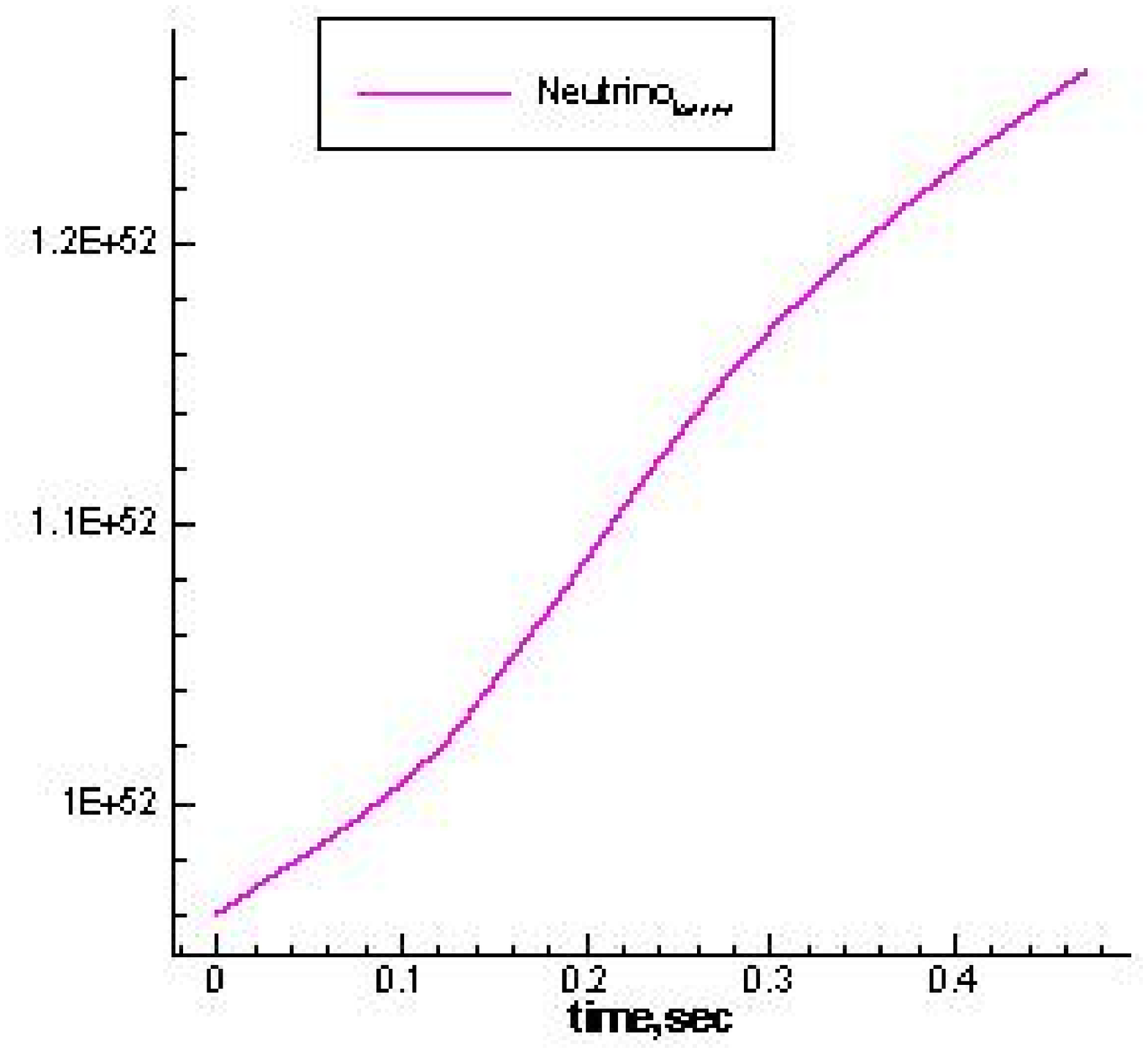,width=3in}
\psfig{figure=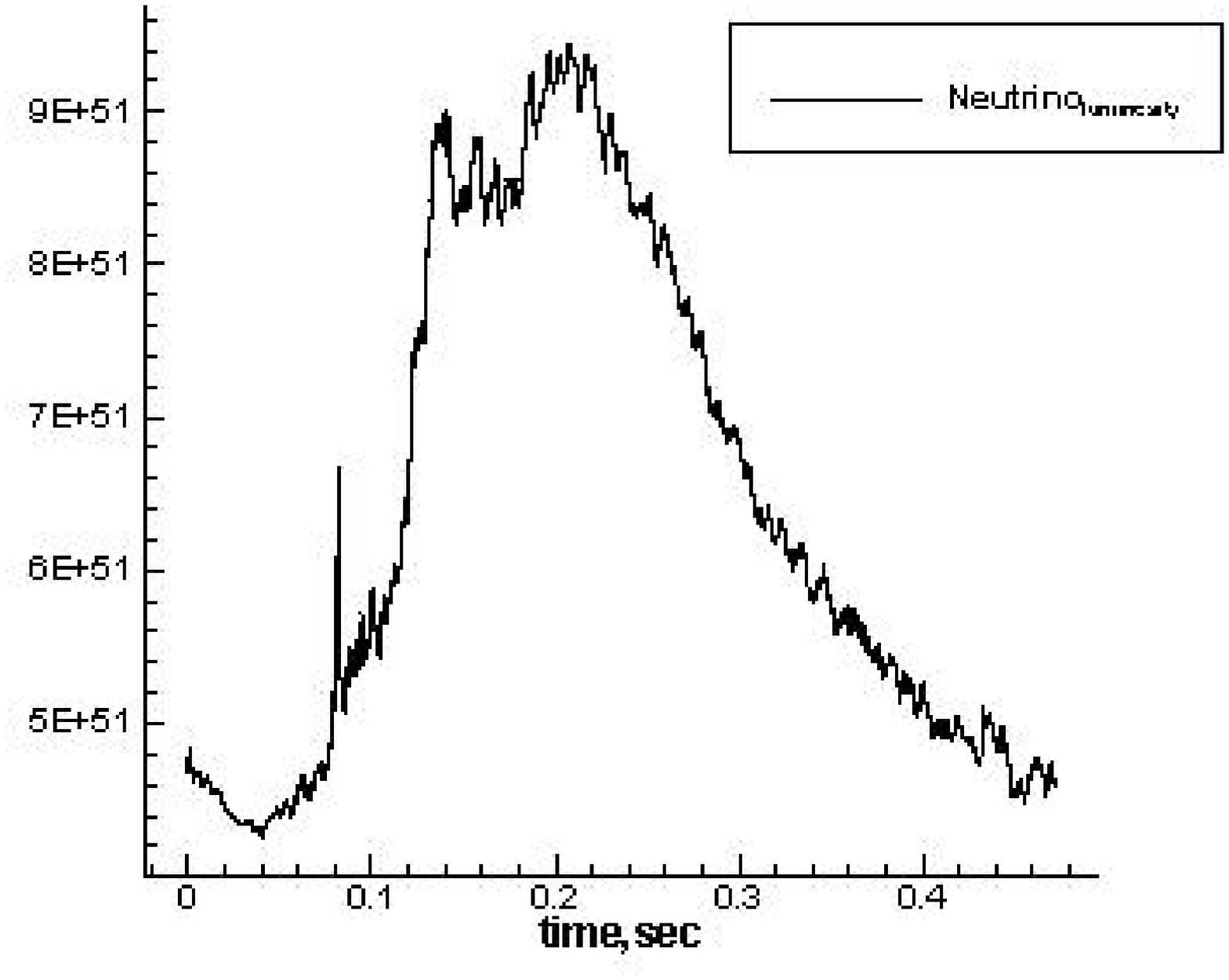,width=3in}}
\caption
{{\small
The integral neutrino losses ({\bf right}),
and neutrino luminosity $\int\limits_0^{M_{core}} f(\rho,T) dm$ ({\bf left}),
as a function of time during the magnetorotational explosion, from \cite{ard05}.}}
\label{nuloss}
\end{figure}

\begin{figure}
\centerline{\psfig{figure=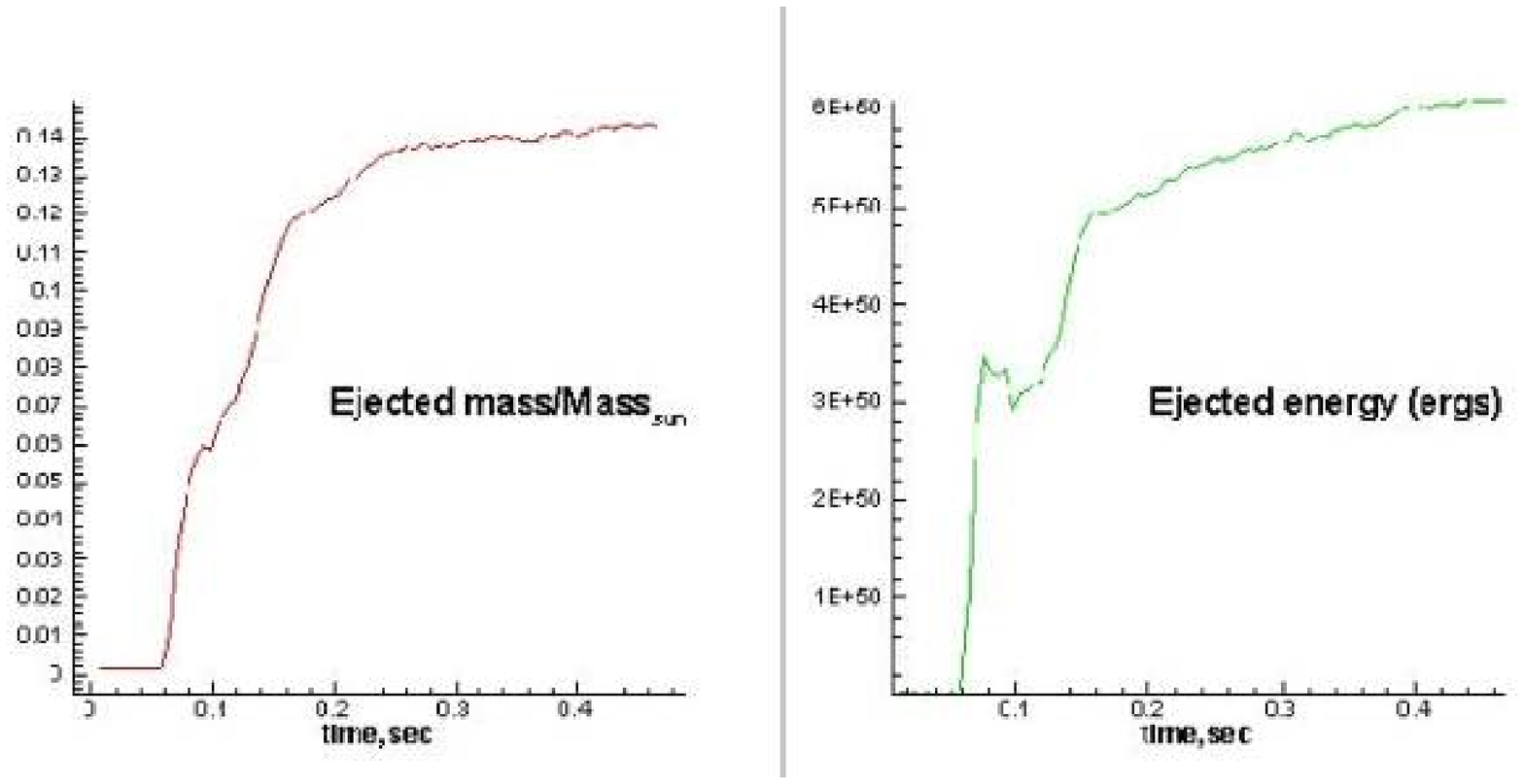,width=6in}}
\caption
{{\small
Time dependence of the ejected mass in ($M_\odot$), and ejected energy
during the magnetorotational explosion with a quadrupole-like field. The
particle is considered  as "ejected"  if its kinetic energy is greater than its potential energy
and its velocity vector is directed from the center, from \cite{ard05}.}}
\label{eject}
\end{figure}

Almost all gravitational energy, transforming into heat during
the collapse, is carried away by weakly interacting neutrino.
Dependence of the neutrino luminosity, and integral neutrino losses on time
are represented in Fig.\ref{nuloss}.
A mass "particle" is considered  "ejected" if its kinetic energy
is greater than its potential energy. Time dependence of the
ejected mass in ($M_\odot$), and the ejected energy
during the magnetorotational explosion with a quadrupole-like field
are represented in Fig.\ref{eject}.
The total energy ejected in the kinetic form is equal to $0.6\cdot 10^{51}$ erg,
and the total ejected mass is equal to $0.14M_\odot$.

\section{Magnetorotational instability}

Magnetorotational instability (MRI)
leads to exponential growth of magnetic fields.
Different types of MRI  have been studied in
\cite{dun58},\cite{vel59},\cite{bh98},\cite{spru02}. MRI starts to
develop when
the ratio of the toroidal to the poloidal magnetic energies is becoming large.
In 1-D calculations MRI is absent because of a restricted degree of freedom.
Therefore the time of the MR explosion is increasing with $\alpha$ as
$t_{\rm expl} \sim \frac{1}{\sqrt{\alpha}}$,
$\alpha=\frac{E_{mag0}}{E_{grav0}}$. Due to the  development of MRI the time of
the MR explosion depends on $\alpha$ much weaker.
The ratio of two magnetic energies
is changing with time almost with the same speed for all $\alpha$, so MRI
starts almost at the same time. The MR explosion happens when the magnetic energy
is becoming comparable to the internal energy, at least in some parts of the star.
While the starting magnetic energy linearly depends on $\alpha$, and MRI leads
to exponential growth of the magnetic energy, the total time of MRE in 2-D is growing
{\bf logarithmically} with decreasing of $\alpha$,
$t_{expl} \sim -\log{\alpha}$. These dependencies are seen
clearly  from 1-D and 2-D calculations with different $\alpha$ giving the following explosion
times $t_{\rm expl}$ (in arbitrary units):

\begin{equation}
\alpha=0.01,\,\, t_{\rm expl}=10, \quad
\alpha=10^{-12},\,\, t_{\rm expl}=10^6 \quad {\rm in\,\,\, 1-D}
\end{equation}
$$
\alpha=10^{-6},\,\, t_{\rm expl}\sim 6, \quad
\alpha=10^{-12},\,\, t_{\rm expl}\sim 12 \quad {\rm in \,\,\,2-D}
$$
\begin{figure}
\centerline{\psfig{figure=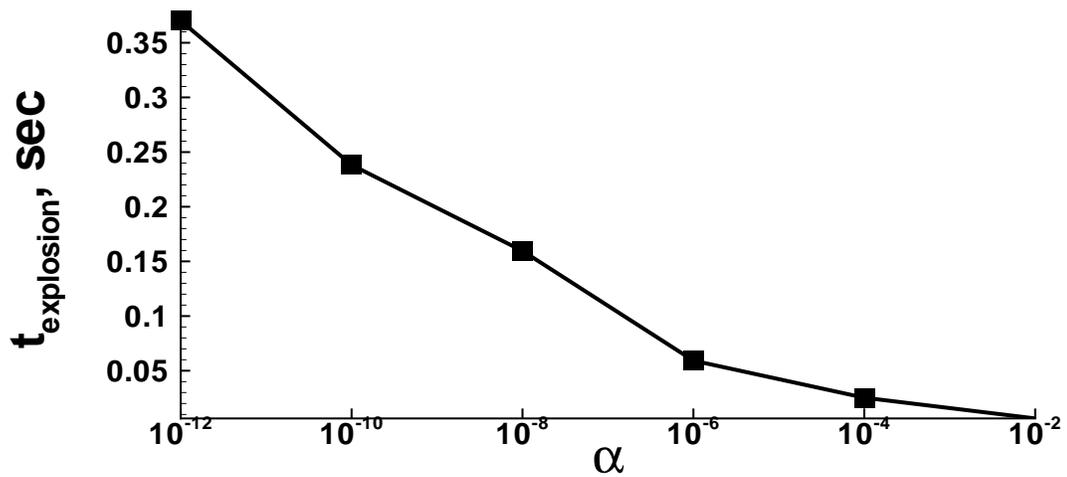,width=6in}}
\caption
{{\small
Duration of the explosion time on $\alpha$ in presence of MRI, from \cite{mois05}}}
\end{figure}

\subsection{Toy model of the MRI development}

\begin{figure}
\centerline{\psfig{figure=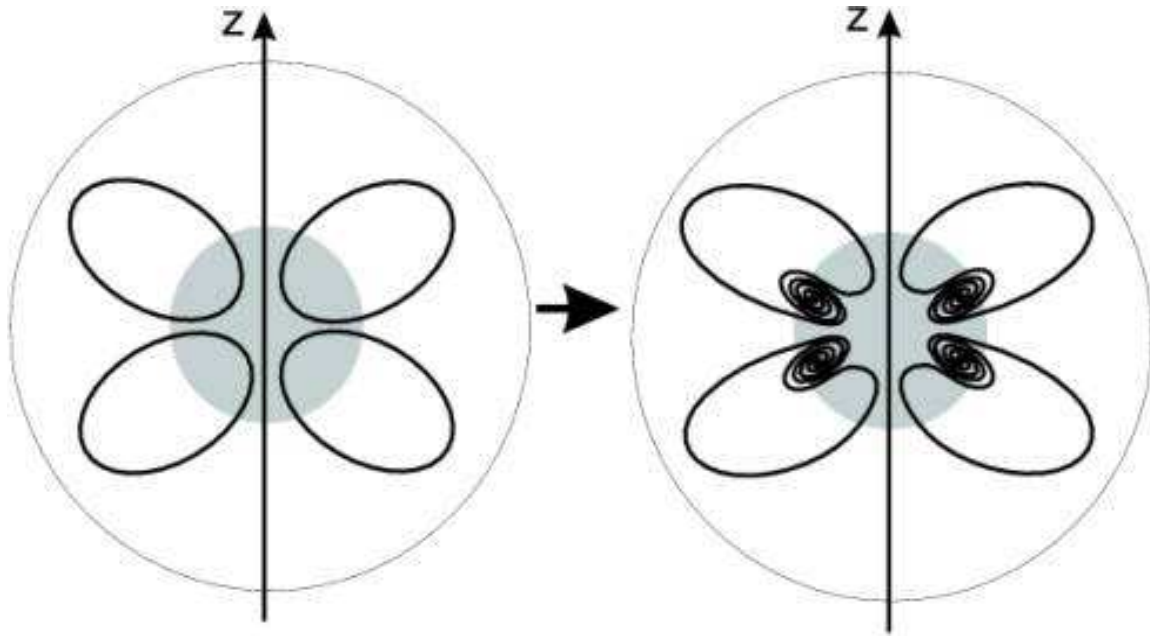,width=6in}}
\caption
{{\small
Qualitative picture of the development of the MRI in 2-D.}}
\label{mri_toy}
\end{figure}

Toy model of the MRI development shows an exponential growth of
the magnetic fields at initial stages of MRI development in 2-D. Initially
MRI leads to formation of multiple poloidal differentially rotating vortexes.
Angular velocity of vortexes is growing (linearly) with a growth of  $H_\varphi$.
\begin{equation}\label{mri1}
\frac{{\mathrm d} H_\varphi}{{\mathrm d} t}=
H_r\left( r\frac{{\mathrm d} \Omega}{{\mathrm d}r}\right).
\end{equation}
The right-hand side is constant at the initial stage of the process.
When the toroidal field reaches its critical value $H_{\varphi}^*$,
the MRI instability starts to develop. As follows from our
calculations, the critical value corresponds to the ratio between total
over the star toroidal and
poloidal magnetic energies  $\sim 1-3$. A local value of this critical ratio
is much larger, reaching at the beginning of MRI
$\sim 10^4$ in the regions of the maximal
growth of the toroidal magnetic field, what in time roughly corresponds to
about 100 rotations of the neutron star core.
These regions serve as germs for subsequent
development of MRI in other parts of the star.
The appearance of MRI is characterized by formation of multiple {\it
poloidal} differentially rotating vortexes, which twist the
initial poloidal field leading to its amplification according to
\begin{equation}\label{mri2}
  \frac{{\mathrm d}H_r}{{\mathrm d} t}=
  H_{r 0}\left(\frac{{\mathrm d} \omega_v}{{\mathrm d} l} l \right),
\end{equation}
where $l$ is the coordinate, directed along the vortex radius,
$\omega_v$ is the angular velocity of the poloidal vortex.
Qualitatively the poloidal field amplification due to the vortexes
induced by MRI is shown in the Fig. \ref{mri_toy}.
The enhanced poloidal field immediately starts to take part in the
toroidal field amplification according to (\ref{mri1}). With
further growing of $H_\varphi$ the poloidal vortex speed
increases. Our calculations give the values of $\omega_v=0.0132\>
{\mathrm s^{-1}}$ at $|H_\varphi|=2.46 \cdot 10^{15}{\mathrm G}$
corresponding to $t=0.041 \>{\mathrm s}$ and $\omega_v=0.052\>
{\mathrm s^{-1}}$ at $|H_\varphi|=4.25\cdot 10^{15}{\mathrm G} $
corresponding to $t=0.052 \>{\mathrm s^{-1}}$ for the same
Lagrangian particle. In general we may approximate the value in
brackets (\ref{mri2}) by linear function on the value
$(H_\varphi-H^*_\varphi )$ as
\begin{equation}\label{mri3}
  \left(\frac{{\mathrm d} \omega_v}{{\mathrm d} l} l \right)=
  \alpha (H_\varphi-H^*_\varphi ).
\end{equation}
\begin{equation}\label{mri4}
  \frac{{\mathrm d}^2}{{\mathrm d} t^2}\left(
  H_\varphi - H^*_{\varphi }\right)=
  A H_{r 0} \alpha (H_\varphi - H^*_{\varphi }),
\end{equation}
\begin{eqnarray*}
  H_\varphi=H^*_{\varphi } + H_{r0} e^{\sqrt{A\alpha H_{r0}}(t-t^*)}, \\
  H_{r}=H_{r0}+\frac{H^{3/2}_{r0} \alpha^{1/2}}
  {\sqrt{A}}\left( e^{\sqrt{A\alpha H_{r0}}(t-t^*)} -1 \right),
\end{eqnarray*}

\section{Jet formation  in MRE}

\begin{figure}
\centerline{\psfig{figure=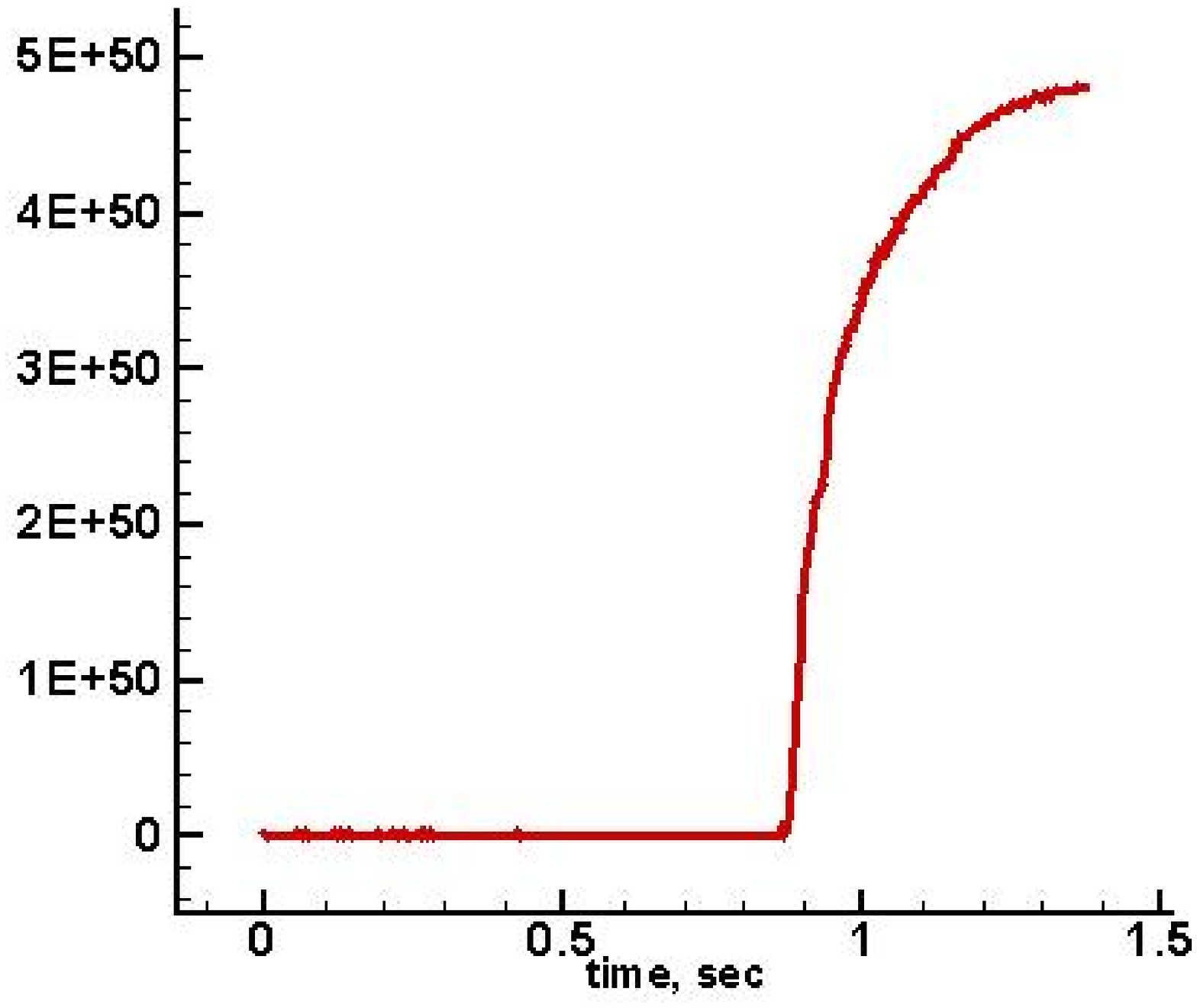,width=3in}
  \psfig{figure=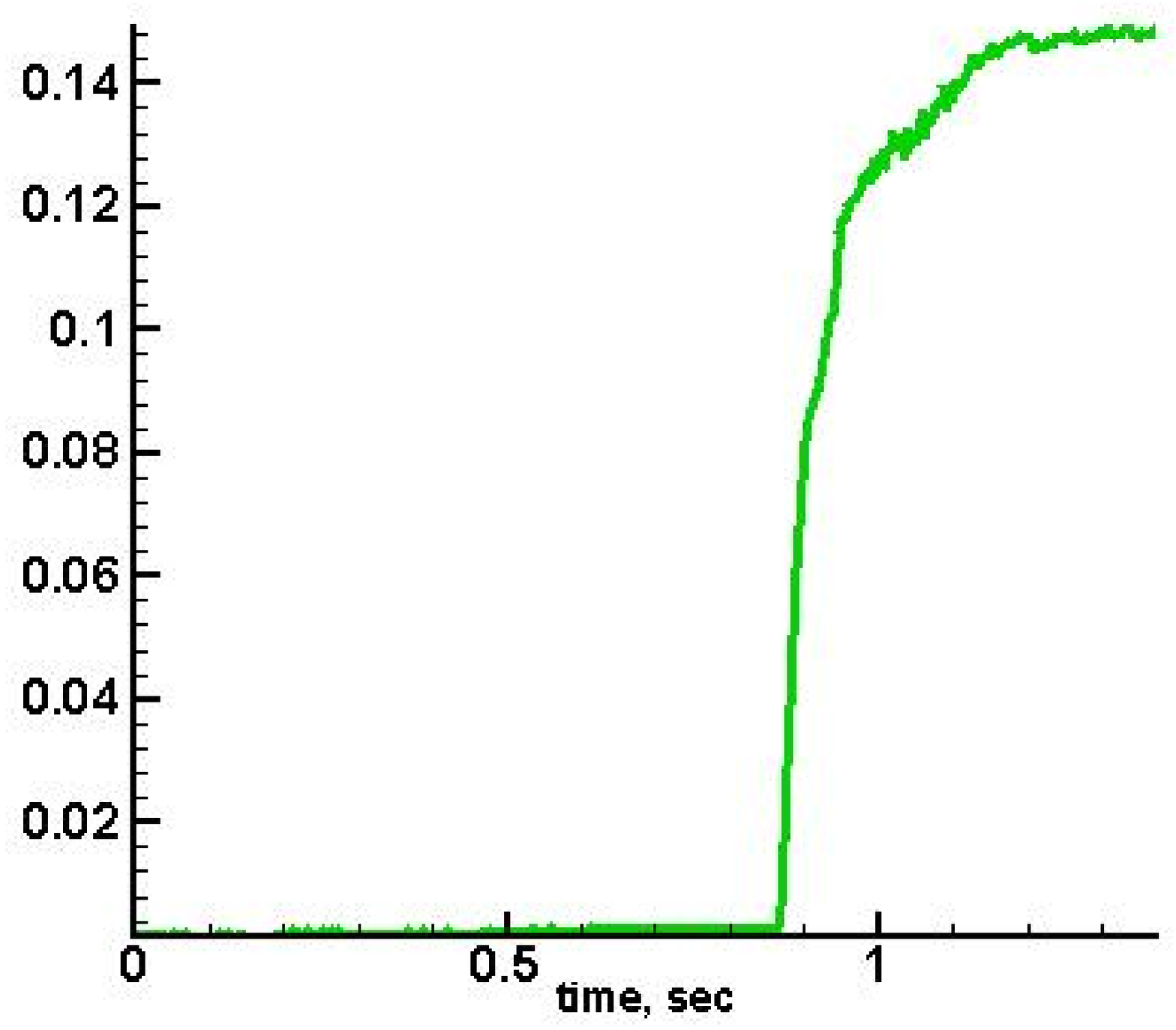,width=3in}}
\caption
{{\small
({\bf Left}) Time dependence of the ejected energy
during the magnetorotational explosion with a dipole-like field.
({\bf right}) Time dependence of the ejected mass
during the magnetorotational explosion with a dipole-like field,
from \cite{mois05}.
}}
\label{dip1}
\end{figure}

Jet formation  in MRE happens when the
initial magnetic field is of a dipole-like structure.
2-D calculations with the initial dipole-like magnetic field gave almost the same
values of the energy of explosion $\sim 0.5 \cdot 10^{51}$ erg,
 and ejected mass $\approx 0.14M_{\odot}$ (see Figs.\ref{dip1}),
but the outburst was slightly collimated
along the rotational axis \cite{mois05}.

\subsection{Violation of mirror symmetry of magnetic field}

\begin{figure}
\centerline{\psfig{figure=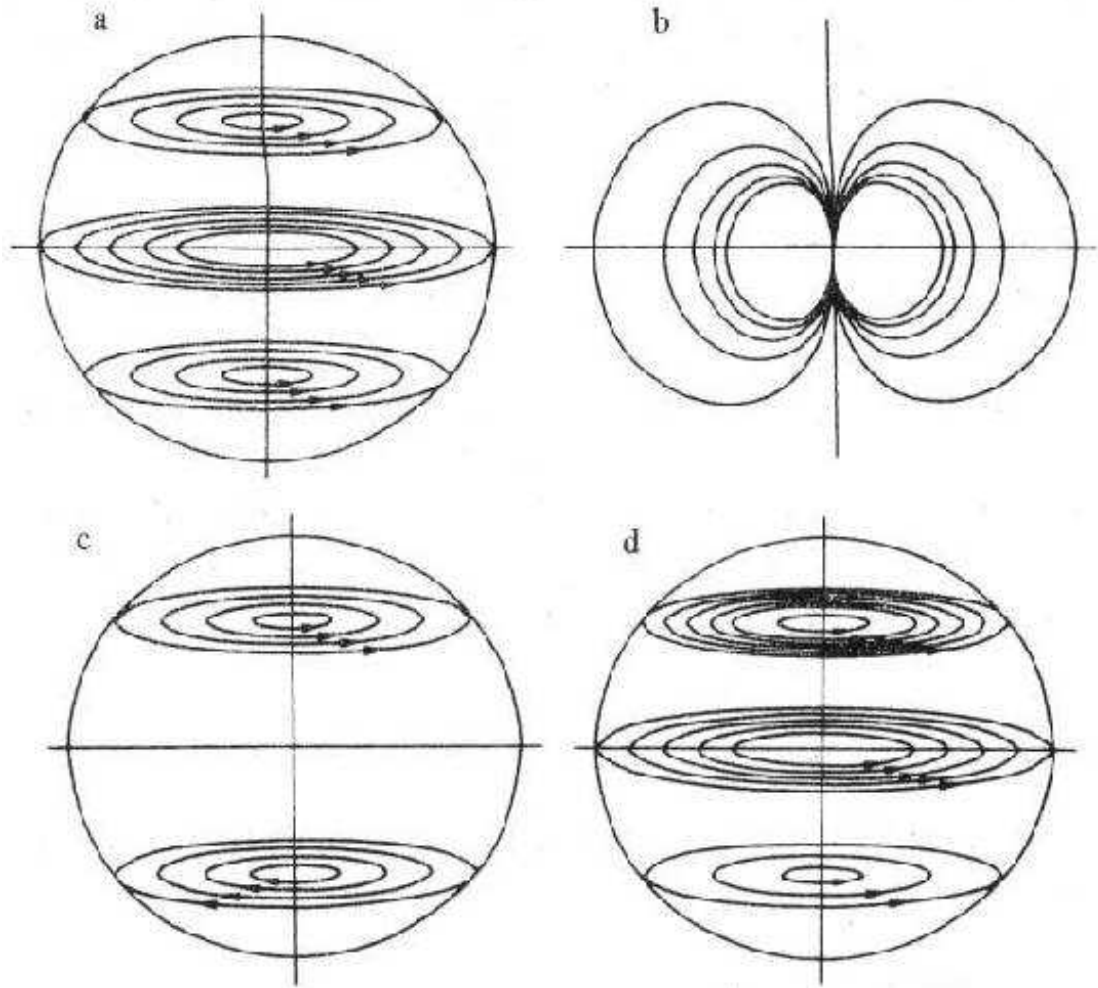,width=6in}}
\caption
{{\small
Loss of a mirror symmetry of the magnetic field in the differentially
rotating star, from \cite{mois92}.}}
\label{mirror}
\end{figure}

There is a violation of the mirror symmetry of magnetic field
when two components dipole + quadrupole are present from the beginning
\cite{lov92}. Another possibility of violation of the mirror symmetry of
magnetic field was considered in \cite{mois92}. If we start from axially symmetric
configuration with a mixture of poloidal and toroidal fields with different
parity, than any exchange between these two components will lead to such violation.
For example, if we start from the dipole-like poloidal field (negative parity), and
symmetric toroidal field with a positive parity, than twisting of the dipole magnetic field
by the differential rotation leads to creation of additional toroidal field with a negative
parity, Therefore the total resulting toroidal field will become asymmetric
relative to the equatorial plane, see Fig.\ref{mirror}.

The magnetorotational explosion with a mirror asymmetric field
will be always asymmetrical.
A kick velocity due to such asymmetry of the
magnetic field could be up to $\sim 300$ km/sec \cite{mois92}.
The interaction of the neutrino with asymmetric magnetic field could
increase the kick velocity up to $\sim 1000$ km/sec, giving a possibility to
explain the origin of velocities of the most rapidly moving pulsars \cite{bk93}.

\section{Conclusions}

\indent
\indent
1. In the magnetorotational explosion (MRE) the efficiency of transformation
of rotational energy into the energy of explosion is 10\%.
This is enough for producing core – collapse SN  from
rapidly rotating magnetized neutron star.

2. Development of magneto-rotational instability strongly
accelerates MRE, at lower values of the initial magnetic fields.

3. The new born neutron star has inside a large
(about $10^{14}$ Gauss) chaotic magnetic field.

4.   Jet formation is possible for dipole-like initial topology of the field, what may have a
possible relation to cosmic gamma-ray bursts; equatorial ejection happens
at prevailing of the quadrupole-like component.

5.    Braking of the equatorial symmetry happens in differentially
rotating star with toroidal and poloidal components of an opposite symmetry;
it may explain formation of the rapidly moving pulsars and one-side jet formation.

\smallskip

{\bf Acknowledgement.} G.S.B.-K. is grateful to the organizers of the
conference for support and hospitality. G.S.B.-K. and S.G.M. Thanks RFBR for the
partial support in the frame of the grant 05-02-17697.

\end{document}